\documentclass[a4paper,11pt]{article}
\usepackage{jheppub}

\usepackage{amsmath,amssymb,amsthm,mathtools}
\usepackage{graphicx}
\usepackage{booktabs}
\usepackage{subcaption}
\usepackage{xcolor}
\usepackage{mathrsfs}
\usepackage{microtype}
\usepackage{amsthm}

\newcommand{\Veff}{\mathcal{V}}
\newcommand{\cA}{\mathcal{A}}
\newcommand{\cL}{\mathcal{L}}
\newcommand{\GN}{G_{\!N}}
\newcommand{\vE}{v_{E}}
\newcommand{\zs}{z_{c}}
\newcommand{\zt}{z_{t}}
\newcommand{\zh}{z_{h}}
\newcommand{\dd}{\mathrm{d}}
\newcommand{\PV}{\mathrm{P.V.}\!}
\newcommand{\Loss}{\mathscr{L}}

\title{Deep learning spacetime from entanglement dynamics}

\author[a,b]{Hyun-Sik Jeong,}
\author[c,d]{Keun-Young Kim,}
\author[c]{Gaya Yun}
\author[c]{and Kwan Yun}

\emailAdd{hyunsik.jeong@apctp.org}
\emailAdd{fortoe@gist.ac.kr}
\emailAdd{gayayun121@gm.gist.ac.kr}
\emailAdd{ludibriphy70@gm.gist.ac.kr}

\preprint{\texttt{APCTP Pre2026 - 017}}

\affiliation[a]{Asia Pacific Center for Theoretical Physics, Pohang 37673, Korea}
\affiliation[b]{Department of Physics, Pohang University of Science and Technology, Pohang 37673, Korea}
\affiliation[c]{Department of Physics and Photon Science, Gwangju Institute of Science and Technology, \\ 123 Cheomdan-gwagiro, Gwangju 61005, Korea}
\affiliation[d]{Research Center for Photon Science Technology, Gwangju Institute of Science and Technology, \\ 123 Cheomdan-gwagiro, Gwangju 61005, Korea}

\abstract{
In the eternal AdS black hole, the entanglement of the thermofield double state is geometrically realised by the Einstein--Rosen bridge, in line with the ER=EPR conjecture. Hartman--Maldacena surfaces threading this bridge pass behind the event horizon, so the time-dependent entanglement entropy $S(t)$ of a pair of half-lines, one on each boundary, probes the black hole interior. We turn this relation around and ask how much of the bulk spacetime can be reconstructed from $S(t)$, using a neural ordinary differential equation that learns the metric directly from the entanglement data. We show that $S(t)$ depends on the geometry only through a single function, the local surface gravity expressed in terms of the timelike Killing norm. This is enough to fix a metric with one unknown function: for charged BTZ black holes, the network reconstructs it throughout the region probed by the surfaces, including part of the interior, and the learned metric also yields the Hawking temperature and the location of the critical surface that bounds the accessible interior. With two unknown functions, however, it is not: physically distinct geometries whose two-dimensional $(t,z)$ sectors are related by diffeomorphisms produce the same $S(t)$. Finally, although the null energy condition is not imposed, the network reproduces data obeying the causal bound $v_E\le 1$ on the entanglement velocity, but not data that violate it. Our results thus offer a data-driven perspective on how spacetime, including the black hole interior, emerges from quantum entanglement, and delineate which features of the geometry entanglement alone can determine.
}

\begin{document}
\maketitle

%
\section{Introduction}\label{sec:intro}

The AdS/CFT correspondence connects quantum field theories on the boundary and gravitational theories in a higher-dimensional spacetime~\cite{Maldacena:1997re,Gubser:1998bc,Witten:1998qj}. In particular, the Ryu--Takayanagi (RT) formula and its covariant generalisation, the Hubeny--Rangamani--Takayanagi (HRT) prescription, relate the entanglement entropy of a boundary region to the area of an extremal surface in the bulk~\cite{Ryu:2006bv,Ryu:2006ef,Hubeny:2007xt}. This connection has led to the idea that the structure of quantum entanglement is closely associated with the emergence of bulk spacetime geometry~\cite{VanRaamsdonk:2010pw,Swingle:2009bg}. It also motivates an inverse question: \textit{to what extent can an unknown bulk geometry be reconstructed from entanglement data in the boundary theory?}

Black hole interiors provide an important setting for investigating this question. An eternal AdS black hole is dual to the thermofield double (TFD) state of two decoupled copies of the boundary theory, whose entanglement is geometrically associated with the Einstein--Rosen bridge connecting the two asymptotic regions~\cite{Maldacena:2001kr}, a relation captured by the so-called ER=EPR conjecture~\cite{Maldacena:2013xja}. Hartman and Maldacena (HM) showed that, as the boundary times evolve, an extremal surface connecting the two boundaries can pass through the event horizon and probe the interior of the black hole~\cite{Hartman:2013qma}.\footnote{In two boundary dimensions HM's bulk computation can be checked directly against the field theory: for the eternal BTZ black hole, the geodesic length reproduces exactly the entanglement entropy of the TFD state of two CFT$_2$'s, and when the region consists of half of each boundary the result is fixed entirely by conformal symmetry~\cite{Hartman:2013qma}.} The corresponding entanglement entropy grows linearly at late times, reflecting the continued growth of the bridge. In contrast, static RT surfaces anchored on a single boundary in the usual stationary setup cannot penetrate the event horizon and probe only the exterior geometry~\cite{Hubeny:2012ry,Engelhardt:2013tra}. Time-dependent entanglement therefore offers a way to investigate bulk regions that are inaccessible to such static observables.

The growth of entanglement has also been studied in closely related contexts. In quantum quenches, the spreading of entanglement can be characterised by an entanglement velocity, providing a connection between entanglement dynamics and the propagation of quantum information~\cite{Calabrese:2005in,Liu:2013iza,Liu:2013qca,Mezei:2016zxg}. In holographic theories, this growth is subject to constraints associated with causality~\cite{Hartman:2015apr,Mezei:2016wfz}. For finite boundary regions, competition between extremal surfaces can lead to the saturation of entanglement growth. In related radiating and doubly holographic black hole setups, transitions involving island surfaces play a central role in Page-curve behaviour~\cite{Page:1993wv,Penington:2019npb,Almheiri:2019psf,Almheiri:2019hni,Almheiri:2020cfm,Almheiri:2019psy,Ling:2020laa,Jeong:2023lkc}. In eternal black holes, the Einstein--Rosen bridge can continue to grow even after the entanglement entropy of a finite region has saturated. This observation has motivated proposals relating the black hole interior to quantum computational complexity~\cite{Susskind:2014rva,Stanford:2014jda,Susskind:2014moa}, and the time dependence of holographic complexity, combined with the HM entanglement entropy, has been used to reconstruct the metric inside static black holes whose exterior is known~\cite{Hashimoto:2021umd}. These developments provide a broader motivation for asking which aspects of the interior geometry are encoded in a given boundary observable, and which require additional information.

Bulk reconstruction from boundary observables has been investigated using geodesic lengths, correlation functions, and entanglement entropy~\cite{Porrati:2003na,Bilson:2008ab,Bilson:2010ff,Jokela:2020auu}. For static entanglement entropy, the relation between the size of a boundary region and the turning point of its bulk extremal surface can, under suitable assumptions, be written as an Abel-type integral equation. Bilson's method uses this structure to reconstruct a metric function from entanglement data~\cite{Bilson:2008ab,Bilson:2010ff}. Such approaches demonstrate how boundary entanglement can constrain bulk geometry, but their direct applications are restricted by the chosen metric ansatz and by the region reached by static extremal surfaces. In particular, reconstructing the geometry behind an event horizon requires observables that probe the black hole interior.

A complementary approach to holographic reconstruction employs machine learning. In the AdS/Deep-Learning (AdS/DL) framework introduced in~\cite{Hashimoto:2018ftp,Hashimoto:2018bnb}, the radial evolution of bulk fields is related to a deep neural network, allowing unknown bulk quantities to be inferred from boundary data. Neural ordinary differential equations (neural ODEs) extend this approach by representing the unknown radial functions continuously and integrating the relevant differential equations during training~\cite{Chen:2018neural,Hashimoto:2020jug}. These techniques have been applied to bulk reconstruction using various boundary observables, including optical conductivity, QCD equation of state, fermionic spectral function, and static holographic entanglement entropy~\cite{Ahn:2024gjf,Jeong:2025omu,Hashimoto:2026hfz,Ahn:2024lkk}. In particular,~\cite{Ahn:2024lkk} used neural ODEs to reconstruct black hole metric functions from static entanglement data. This motivates extending the inverse problem to time-dependent entanglement, for which the extremal surfaces can reach beyond the horizon.

However, reconstructing the bulk metric from HM entanglement dynamics is not a straightforward extension of the static problem. In the static case, the turning point can be related directly to the boundary entanglement data, which makes Abel-type inversion possible in suitable backgrounds. For HM surfaces, the turning point lies behind the horizon and depends on the unknown metric, while the conventional integral representation involves a principal-value prescription at the horizon. Consequently, the static reconstruction procedure cannot be applied directly to real-time HM data (see section~\ref{sec:abel}). An alternative is the differential description of HM surfaces employed by~\cite{Jeong:2023lkc}, which allows their evolution across the horizon to be treated without explicit principal-value integration. Combined with a neural ODE framework, this provides a natural starting point for studying the corresponding inverse problem. Beyond the numerical reconstruction itself, an important question is how much geometric information the time-dependent entanglement entropy can determine.

In this work, we investigate the reconstruction of three-dimensional black hole spacetime from the time-dependent entanglement entropy $S(t)$ of a TFD state. We consider HM surfaces associated with a pair of half-lines, one on each boundary, and employ their differential equations within a neural ODE framework. We use BTZ and charged BTZ black holes as representative backgrounds and study the reconstruction of the metric functions from the boundary dynamical entanglement data. We also examine whether the metric is uniquely determined by the data and how physical conditions constrain the reconstruction.
We find that the half-space entanglement dynamics depends on the bulk metric only through a single function, the local surface gravity as a function of the Killing norm: with one unknown metric function the network reconstructs the geometry over the entire region probed by the surfaces, including part of the interior, while with two unknown functions we construct a family of distinct geometries with identical $S(t)$, and the network indeed finds such geometries.
We also show that the null energy condition bounds the rate of entanglement growth by the thermal entropy density and the speed of light, and that data exceeding this bound are not reproduced by the network.

The paper is organised as follows. Section~\ref{sec:hm} reviews the integral and differential descriptions of HM surfaces and discusses their entanglement dynamics in BTZ and charged BTZ backgrounds. Section~\ref{sec:recon} introduces the inverse problem and the neural ODE framework, followed by the analysis of possible non-uniqueness and constraints from causality. We conclude in section~\ref{sec:conclusion}. Appendix~\ref{app:cft} reviews the corresponding CFT$_2$ calculation and appendix~\ref{app:gen} presents the HM equations in general dimensions.

%
\section{Holographic dynamics of entanglement entropy}\label{sec:hm}

\begin{figure}[t]
\centering
\includegraphics[width=1.0\textwidth]{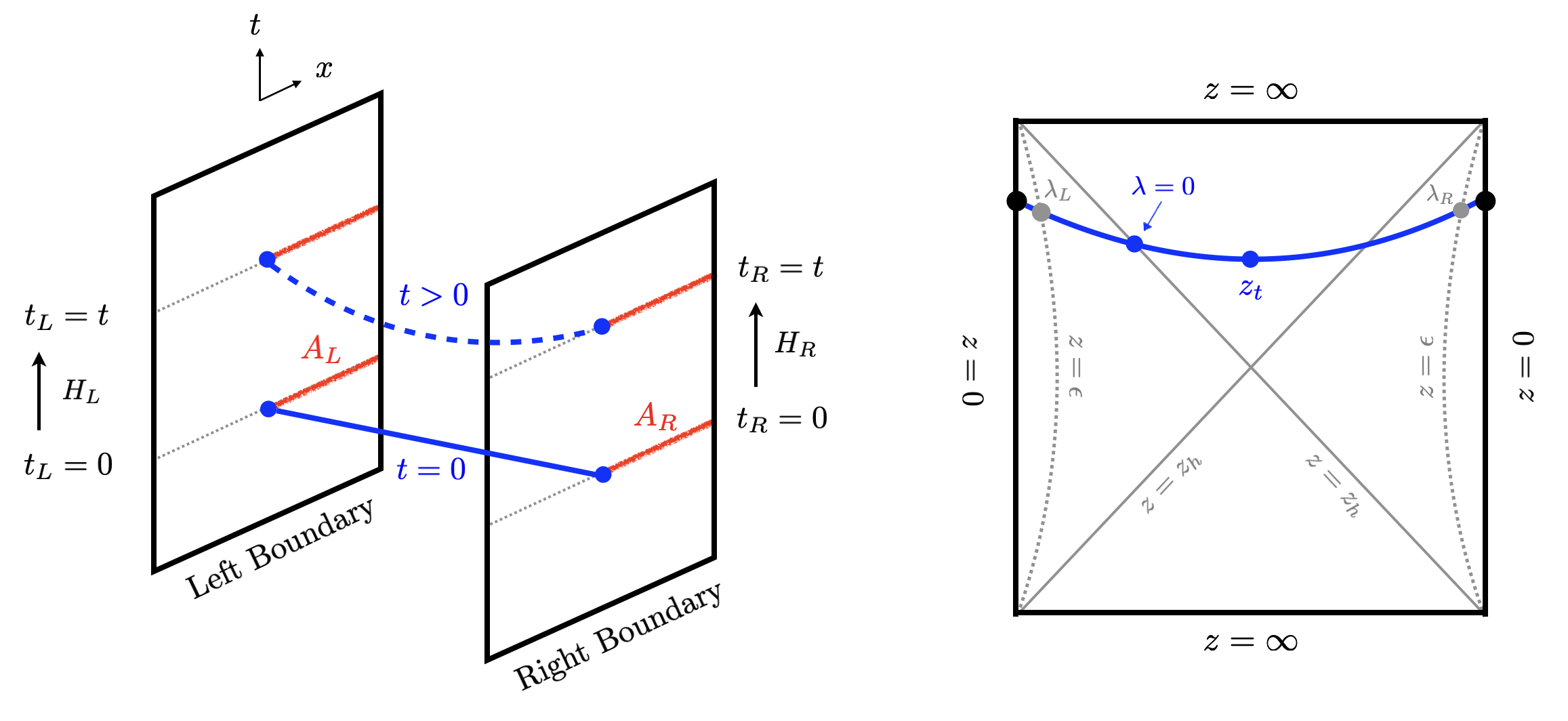}
\caption{Setup of the half-space entanglement dynamics. \textbf{Left:} The two asymptotic boundaries of the TFD state. The subsystem $A=A_L\cup A_R$ consists of the half-lines $x>0$ (red) on the left and right boundaries; the complementary half-lines $x<0$ are dotted. Both copies are evolved with $H_L+H_R$, so that both boundary times advance, $t_L=t_R=t$; the configurations at $t=0$ (solid blue) and at $t>0$ (dashed blue) are shown. The entangling surface consists of the two points $x=0$ (blue dots), which the HM surface (blue) connects through the Einstein--Rosen bridge; its area, and hence $S(t)={\cA(t)}/{4\GN}$, grows with $t$. \textbf{Right:} Schematic Penrose diagram of the $x=0$ slice of the two-sided black hole, with the HM surface at boundary times $t_L$ and $t_R$. The surface crosses the horizon ($z=z_h$) and reaches a turning point $z_t$ behind the horizon. The dashed gray curves indicate the UV cutoff $z=\epsilon$, where the surface is terminated at $\lambda_L$ and $\lambda_R$. The point $\lambda=0$ on the left future horizon serves as the initial point for numerical integration.}
\label{fig:penrose}
\end{figure}

We consider the following asymptotically AdS$_3$ metric, setting the AdS radius $L=1$:
\begin{equation}\label{eq:metric}
\dd s^2=\frac{1}{z^2}\left(-f(z)e^{-\chi(z)}\dd t^2+\frac{\dd z^2}{f(z)}+\dd x^2\right).
\end{equation}
The boundary is at $z=0$, where $f(0)=1$ and $\chi(0)=0$ so that $t$ is the boundary time, and the outer horizon is at $z=\zh$, $f(\zh)=0$.
The Hawking temperature and the thermal entropy density are
\begin{equation}\label{eq:thermo}
T=\frac{|f'(\zh)|\,e^{-\chi(\zh)/2}}{4\pi}\equiv\frac{\kappa_h}{2\pi}\,,\qquad s=\frac{1}{4\GN\zh}\,.
\end{equation}
The maximally extended geometry is dual to the TFD state~\cite{Maldacena:2001kr}
\begin{equation}
|\mathrm{TFD}(t)\rangle=\mathcal{Z}^{-1/2}\sum_n e^{-\beta E_n/2}\,e^{-2iE_nt}\,|n\rangle_L|n\rangle_R\,,
\end{equation}
where $\beta=1/T$ is the inverse temperature, $E_n$ are the energies of a single copy and $\mathcal{Z}$ is its partition function.
We evolve this state with $H_L+H_R$, so that both boundary times advance, $t_L=t_R\equiv t$~\cite{Hartman:2013qma}.
In the results presented below, we use the total boundary time $\Delta t\equiv t_L+t_R=2t$, and express the entanglement entropy $S(t)$ in units with $\GN=1$.
We take the subsystem $A=A_L\cup A_R$ to be the union of the half-lines $x>0$ on the two boundaries.
Its entangling surface consists of the two points $x=0$, and by the reflection symmetry $x\to-x$ the HRT surface is the spacelike geodesic at $x=0$ that connects the two boundaries through the Einstein--Rosen bridge: the HM surface (figure~\ref{fig:penrose}).
Since each boundary contains a single endpoint, no other surface is homologous to $A$, and
\begin{equation}
S(t)=\frac{\cA(t)}{4\GN}\,,
\end{equation}
with $\cA$ the area of the HM surface.

%
\subsection{Hartman--Maldacena integral formalism: a review}\label{sec:integral}

By the time-reflection symmetry of the TFD state, the HM surface is symmetric about its midpoint, a turning point $z=\zt$ that lies behind the future horizon.
Parametrising one half of the surface by $t(z)$, its area is
\begin{equation}\label{eq:lengthfunc}
\cA=2\int\dd z\;\frac{1}{z}\sqrt{\frac{1}{f}-f e^{-\chi}\,t'(z)^2}\,.
\end{equation}
Time-translation invariance implies a conserved quantity, the energy
\begin{equation}\label{eq:Edef}
E=\frac{f e^{-\chi}\,t'}{z\sqrt{f^{-1}-f e^{-\chi}t'^2}}\,,
\end{equation}
which we take to be positive for $t>0$.
Solving for $t'$ gives
\begin{equation}\label{eq:tprime}
t'(z)=\frac{E\,e^{\chi/2}}{f\sqrt{E^2-\Veff(z)}}\,,\qquad
\Veff(z)\equiv-\frac{f(z)\,e^{-\chi(z)}}{z^2}\,,
\end{equation}
where we used $f+E^2z^2e^{\chi}=z^2e^{\chi}\left(E^2-\Veff\right)$.
In terms of the proper length $\lambda$ along the surface, the same statement reads
\begin{equation}\label{eq:scattering}
\left(\frac{\dd z}{\dd\lambda}\right)^2=z^4e^{\chi}\big(E^2-\Veff(z)\big)\,,
\end{equation}
which is the energy equation of a particle with energy $E^2$ moving in the potential $\Veff(z)$, with a position-dependent mass~\cite{Hartman:2013qma}.
Note that $\Veff=g_{tt}$ is the norm of the timelike Killing vector $\partial_t$.
It is negative outside the horizon, vanishes at $z=\zh$ and is positive inside; it diverges as $-1/z^2$ near the boundary.
A surface with energy $E$ enters the black hole and turns around at $\zt$, where
\begin{equation}\label{eq:turning}
E^2=\Veff(\zt)\,.
\end{equation}
For the geometries of interest, $\Veff$ has a maximum $\Veff(\zs)=E_c^2$ at a critical surface $\zs$ inside the horizon, $\zh<\zs$, defined by $\Veff'(\zs)=0$, i.e.,
\begin{equation}\label{eq:zstar}
\zs\big(f'(\zs)-f(\zs)\chi'(\zs)\big)=2f(\zs)\,.
\end{equation}
As $E$ increases from $0$ to $E_c$, the turning point moves from $\zh$ to $\zs$, and the boundary time grows from $0$ to infinity.
The critical surface is therefore the maximal depth $z_{\max}=\zs$ that HM surfaces can reach at late time.

Using \eqref{eq:lengthfunc} and \eqref{eq:tprime}, the area and the boundary time are given by
\begin{align}
\cA(E)&=2\int_{\epsilon}^{\zt}\dd z\;\frac{e^{-\chi/2}}{z^2\sqrt{E^2-\Veff(z)}}\,,\label{eq:AE}\\
t(E)&=-\,\PV\!\int_0^{\zt}\dd z\;\frac{E\,e^{\chi/2}}{f\sqrt{E^2-\Veff(z)}}\,,\label{eq:tE}
\end{align}
where $\epsilon$ is the UV cutoff, $\cA=2\log(1/\epsilon)+\cA_{\rm reg}$.
The integrand of~\eqref{eq:tE} has a simple pole at the horizon, $t'\simeq e^{\chi(\zh)/2}\big/\big[f'(\zh)(z-\zh)\big]$, because the Schwarzschild time diverges there although the surface itself is perfectly smooth.
The prescription is obtained by continuing $t$ across the future horizon through the ingoing coordinate $t_{\rm in}=t-z_*(z)$, with the tortoise coordinate
\begin{equation}\label{eq:tortoise}
z_*(z)=\int_0^z\dd\tilde z\;\frac{e^{\chi(\tilde z)/2}}{f(\tilde z)}\,,
\end{equation}
which is regular there: $\dd t_{\rm in}/\dd z=t'-e^{\chi/2}/f$ is finite at $\zh$ since both terms have the same residue.
Deforming the contour around the pole produces an imaginary part $\pm i\beta/4$, which is precisely the imaginary shift between the Schwarzschild time coordinates of the exterior and interior regions; the real boundary time is given by the principal value~\cite{Hartman:2013qma,Carmi:2017jqz}.
In practice~\eqref{eq:tE} is evaluated by subtracting the pole explicitly.

The Hamilton--Jacobi relation between the on-shell area~\eqref{eq:AE} and the boundary time~\eqref{eq:tE} gives
\begin{equation}\label{eq:HJ}
\frac{\dd S}{\dd t}=\frac{E}{2\GN}\,,\qquad\text{or}\qquad\frac{\dd\cA}{\dd t}=2E\,.
\end{equation}
Thus the conserved energy of the HM surface is a boundary observable: it is the slope of the entanglement entropy.
We will use this fact repeatedly.

\paragraph{Late-time growth and the entanglement velocity.}
At late times $E\to E_c$ and the surface lingers near $\zs$, so that both $\cA$ and $t$ diverge logarithmically in $E_c-E$.
By~\eqref{eq:HJ}, the entropy grows linearly, $S\simeq E_c \, t/(2\GN)$.
Comparing with the thermal entropy density, the rate of growth defines the entanglement velocity ($v_E$) through $\dd S/\dd t=s\,\vE\,|\partial A|$ with $|\partial A|=2$ the number of entangling points~\cite{Liu:2013iza,Liu:2013qca}:
\begin{equation}\label{eq:vE}
\vE=\zh E_c=\frac{\zh}{\zs}\sqrt{-f(\zs)\,e^{-\chi(\zs)}}\,,
\end{equation}
with $\zs$ given by~\eqref{eq:zstar}.
This is the $d=2$ case of~\eqref{eq:gen_vE} for $(d+1)$-dimensional bulks~\cite{Hartman:2013qma,Liu:2013qca,Mezei:2016zxg}.

%
\subsection{Differential approach to entanglement evolution}\label{sec:differential}

The integral formalism is ideal for the forward problem, but it has two features that could be inconvenient for the inverse problem: the principal value at the horizon, whose position itself depends on the metric, and the integrable singularity at the turning point, whose location~\eqref{eq:turning} must be found by root-finding for every surface.
We now review the differential formalism of~\cite{Jeong:2023lkc}, which avoids both, specialising it to three dimensions and to the metric~\eqref{eq:metric}.

We introduce the Kruskal-like coordinate
\begin{equation}\label{eq:Vcoord}
V=e^{\kappa_h\,t_{\rm in}}=e^{\kappa_h\,(t-z_*(z))}\,,\qquad \kappa_h=2\pi T\,,
\end{equation}
built from the ingoing Eddington--Finkelstein coordinate $t_{\rm in}$ and the tortoise coordinate~\eqref{eq:tortoise}.
The metric becomes
\begin{equation}\label{eq:metricV}
\dd s^2=\frac{1}{z^2}\left(-\frac{f e^{-\chi}}{\kappa_h^2V^2}\,\dd V^2-\frac{2\,e^{-\chi/2}}{\kappa_h\,V}\,\dd V\dd z+\dd x^2\right).
\end{equation}
For $V>0$ these coordinates cover the right exterior ($z<\zh$) and the future interior ($z>\zh$), crossing the future horizon of the right exterior smoothly; for $V<0$ they cover the left exterior.
The surface $V=0$ is the remaining branch of the future horizon (the future horizon of the left exterior), which these coordinates map to the single point $z=\zh$.
Describing the HM surface by $z(\lambda)$ and $V(\lambda)$, its area is $\cA=\int\dd\lambda\,\cL$ with
\begin{equation}\label{eq:L1}
\cL=\frac{1}{z}\left(-\frac{f e^{-\chi}}{\kappa_h^2V^2}\,V'^2-\frac{2\,e^{-\chi/2}}{\kappa_h\,V}\,V'z'\right)^{1/2},
\end{equation}
where a prime denotes $\dd/\dd\lambda$.
Because $\cL$ is reparametrisation invariant, its two Euler--Lagrange equations are not independent; following~\cite{Jeong:2023lkc}, we introduce an auxiliary einbein $\eta(\lambda)$,
\begin{equation}
\cA=\frac12\int\dd\lambda\left(\frac{\cL^2}{\eta}+\eta\right),
\end{equation}
which reproduces~\eqref{eq:L1} on the solution of its equation of motion, $\eta=\cL$; we fix the gauge $\eta=1$.
The equations of motion then consist of the second-order equation
\begin{equation}\label{eq:Veq}
V''=\left[1+\frac{e^{-\chi/2}\left(f'-f\chi'\right)}{2\kappa_h}-\frac{f\,e^{-\chi/2}}{\kappa_h\,z}\right]\frac{V'^2}{V}\,,
\end{equation}
where $f'$ and $\chi'$ now denote $z$-derivatives, together with the constraint
\begin{equation}\label{eq:constraint}
\cL^2=1\,,
\end{equation}
which says that $\lambda$ is the proper length, so that $\cA=\lambda_R-\lambda_L$.

The Killing vector $\partial_t=\kappa_h V\partial_V$ yields the conserved energy
\begin{equation}\label{eq:EV}
E=-\frac{1}{z^2}\left(\frac{f e^{-\chi}}{\kappa_h}\,\frac{V'}{V}+e^{-\chi/2}z'\right),
\end{equation}
and combining~\eqref{eq:EV} with~\eqref{eq:constraint} reproduces~\eqref{eq:scattering}.
The energy in~\eqref{eq:EV} uses the orientation of the proper-length parameter and agrees with~\eqref{eq:Edef} on the branch with $z'<0$.

Instead of~\eqref{eq:Veq}, we take as equations of motion the second-order equation for $z(\lambda)$ that follows by differentiating~\eqref{eq:scattering} with respect to $\lambda$,
\begin{equation}\label{eq:zeq}
z''=\frac12\,\frac{\dd}{\dd z}\Big[z^2f(z)+E^2z^4e^{\chi(z)}\Big]
=zf+\frac12z^2f'+E^2e^{\chi}\Big(2z^3+\frac12z^4\chi'\Big)\,,
\end{equation}
together with the first-order equation for $\log|V|$ that follows from~\eqref{eq:EV} and~\eqref{eq:constraint},
\begin{equation}\label{eq:logV}
\big(\log|V|\big)'=\frac{\kappa_h\,z^2}{E\,z^2-e^{-\chi/2}z'}\,.
\end{equation}
The pair of equations~\eqref{eq:zeq} and~\eqref{eq:logV} is equivalent to the original system~\eqref{eq:Veq} and~\eqref{eq:constraint}, provided that the initial conditions satisfy~\eqref{eq:scattering}, which is preserved by~\eqref{eq:zeq}. Equation~\eqref{eq:zeq} is regular at the turning point, while~\eqref{eq:logV} is regular at the right future horizon but has a simple pole at the left future horizon, where $V=0$.

The HM surface leaves the left exterior through the left future horizon, $V=0$, enters the future interior, turns around at $\zt$, and crosses the right future horizon at a finite positive $V$ before reaching the right boundary.
Following~\cite{Jeong:2023lkc}, we shoot from the point where it crosses $V=0$, which we place at $\lambda=0$, so that $(V,z)=(0,\zh)$ there.
Near this point, the solution is fixed order by order by~\eqref{eq:Veq}, \eqref{eq:constraint} and~\eqref{eq:zeq}:
\begin{align}\label{eq:series}
V(\lambda)&=V_{(1)}\lambda+\left(\frac{1}{\zh}-\frac{f''(\zh)}{2f'(\zh)}+\frac34\,\chi'(\zh)\right)z_{(1)}V_{(1)}\,\lambda^2+\mathcal{O}(\lambda^3)\,,\\
z(\lambda)&=\zh+z_{(1)}\lambda+\frac14\Big(\zh^2f'(\zh)+E^2e^{\chi(\zh)}\big(4\zh^3+\zh^4\chi'(\zh)\big)\Big)\lambda^2+\mathcal{O}(\lambda^3)\,,
\end{align}
which agrees with eq.~(35) of~\cite{Jeong:2023lkc} for $d=2$.
The two shooting parameters have a transparent meaning.
The boost symmetry $V\to cV$ (the Killing flow, which shifts $t_R\to t_R+\kappa_h^{-1}\log c$ and $t_L\to t_L-\kappa_h^{-1}\log c$) allows us to set $V_{(1)}=1$ without loss of generality.
Evaluating~\eqref{eq:EV} at $\lambda=0$ gives
\begin{equation}\label{eq:z1E}
z_{(1)}=e^{\chi(\zh)/2}\,\zh^2\,E\,.
\end{equation}
Hence the only free datum at the horizon is the conserved energy, and, by the Hamilton--Jacobi relation~\eqref{eq:HJ}, the latter is the slope of $S(t)$.
This turns the computation of an HM surface into an initial-value problem whose initial data are fixed by the boundary data.
Explicitly, the initial conditions of~\eqref{eq:zeq} are $z(0)=\zh$ and $z'(0)=z_{(1)}$; in practice we start at $\lambda=\pm\delta$, slightly away from the horizon, with $z(\pm\delta)=\zh\pm z_{(1)}\delta+z_{(2)}\delta^2$ and $z'(\pm\delta)=z_{(1)}\pm2z_{(2)}\delta$, where $z_{(2)}$ is the coefficient of $\lambda^2$ in~\eqref{eq:series}.

Integrating towards $\lambda>0$ and $\lambda<0$ until $z(\lambda_R)=z(\lambda_L)=\epsilon$, the area is $\cA=\lambda_R-\lambda_L$, and the boundary times follow from~\eqref{eq:Vcoord}, $t_R=\kappa_h^{-1}\log V(\lambda_R)+z_*(\epsilon)$ and $t_L=-\kappa_h^{-1}\log|V(\lambda_L)|-z_*(\epsilon)$, with $z_*(\epsilon)\to0$.
For the symmetric configuration $t_L+t_R=2t$,
\begin{equation}\label{eq:tV}
t=\frac{1}{2\kappa_h}\log\left|\frac{V(\lambda_R)}{V(\lambda_L)}\right|,\qquad
4\GN S_{\rm reg}(t)=\lim_{\epsilon\to0}\Big(\lambda_R-\lambda_L-2\log\tfrac{1}{\epsilon}\Big)\,.
\end{equation}
The numerical evaluation of~\eqref{eq:tV} is used for all results in this paper. Note that no reference to the turning point, and no principal value, is needed: the right future horizon is crossed smoothly, the left one is handled by the expansion~\eqref{eq:series}, and the time is read off from $V$ at the two boundaries.

\paragraph{Why the differential form is better suited to the inverse problem.}
Both formalisms compute the same surfaces. For learning the metric, however, the differential form has several advantages.
(i) The HM surface is obtained by repeatedly stepping forward in $\lambda$, and each step uses only the metric functions at the surface's current depth $z(\lambda)$. This is exactly how a neural ODE works: the state is updated layer by layer by the same network, with $\lambda$ acting as a continuous layer index.
(ii) In the integral form, the pole of~\eqref{eq:tE} sits at the zero of $f$, which moves when $f$ is varied; the derivative of the principal value with respect to $f$ near the horizon is numerically delicate.
In the $V$-coordinates the horizon is a regular point of the evolution, and $\kappa_h$ enters only as a parameter.
(iii) The turning point is a regular point of the evolution equations~\eqref{eq:zeq} and~\eqref{eq:logV}, so no root-finding of~\eqref{eq:turning} and no integrable singularity are involved, and the same code applies to any $(f,\chi)$ for which the surfaces exist.
(iv) By~\eqref{eq:z1E} and~\eqref{eq:HJ}, the initial data are boundary observables, so each surface is labelled directly by the data.

%
\subsection{Applications and case studies}\label{sec:cases}

Before turning to the inverse problem, we use the differential approach to study the entanglement dynamics that will serve as input entanglement data in section~\ref{sec:recon}.
We set $\zh=1$ in numerical results.

\subsubsection{Analytic study: neutral BTZ black holes and CFT\texorpdfstring{$_2$}{2}}\label{sec:btz}

For the BTZ black hole~\cite{Banados:1992wn}, $f=1-z^2/\zh^2$ and $\chi=0$, so that $\kappa_h=1/\zh$ and $\beta=2\pi\zh$.
The potential $\Veff=\zh^{-2}-z^{-2}$ increases monotonically towards the singularity, so the critical surface sits at $\zs=\infty$ and $E_c=1/\zh$, i.e.\ $\vE=1$ by~\eqref{eq:vE}.
The equations~\eqref{eq:zeq} and~\eqref{eq:logV} can be solved in closed form.
With $\zh=1$ and $V_{(1)}=1$,
\begin{equation}\label{eq:btzsol}
z(\lambda)=\frac{\cosh t}{\cosh(\lambda-t)}\,,\qquad
V(\lambda)=\frac{2\cosh t\,\sinh(\lambda/2)}{\cosh(\lambda/2-t)}\,,\qquad
E=\tanh t\,,
\end{equation}
which satisfies~\eqref{eq:series} and~\eqref{eq:z1E} with $z_{(1)}=\tanh t$.
The turning point is reached at $\lambda=t$, where $\zt=\cosh t$, and the right future horizon is crossed at $\lambda=2t$.
At the cutoff, $\lambda_{R}\simeq t+\log(2\cosh t/\epsilon)$ and $\lambda_L\simeq t-\log(2\cosh t/\epsilon)$, while $V(\lambda_R)\to2\cosh t\,e^{t}$ and $V(\lambda_L)\to-2\cosh t\,e^{-t}$.
Equation~\eqref{eq:tV} then confirms that the boundary time is $t$, and restoring $\zh$,
\begin{equation}\label{eq:btzS}
S(t)=\frac{1}{2\GN}\log\left(\frac{2\zh}{\epsilon}\cosh\frac{t}{\zh}\right)
=\frac{c}{3}\log\left(\frac{\beta}{\pi\epsilon}\cosh\frac{2\pi t}{\beta}\right),
\end{equation}
where we used the Brown--Henneaux central charge $c=3/(2\GN)$~\cite{Brown:1986nw}.
This coincides with the entanglement entropy of two half-lines in the TFD state of any two-dimensional CFT, computed from the two-point function of twist operators on the thermal cylinder (appendix~\ref{app:cft})~\cite{Hartman:2013qma,Calabrese:2004eu,Calabrese:2005in}.

\begin{figure}[t]
\centering
\includegraphics[width=\textwidth]{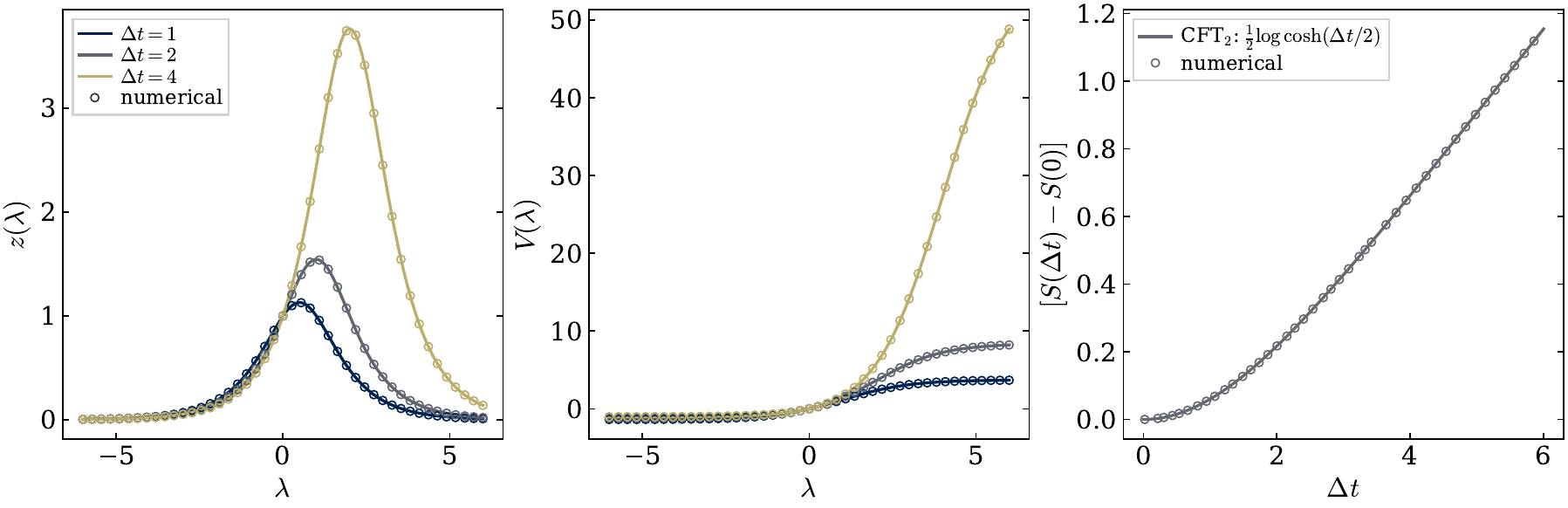}
\caption{BTZ black hole ($\zh=1$). \textbf{Left and middle:} the analytic solution~\eqref{eq:btzsol} for $z(\lambda)$ and $V(\lambda)$ (lines) compared with the numerical integration of the differential equations from the horizon point $(V,z)=(0,\zh)$ (circles), for $\Delta t=1,2,4$ and $z_h=1$. \textbf{Right:} $S(\Delta t)-S(0) = S_{\rm reg}(\Delta t)-S_{\rm reg}(0)$ ($\GN=1$) from the differential approach (circles) compared with the CFT$_2$ result~\eqref{eq:btzS}, $\tfrac12\log\cosh(\Delta t/2)$ (line).}
\label{fig:btz}
\end{figure}

Figure~\ref{fig:btz} compares the analytic solution~\eqref{eq:btzsol} with the numerical solution of the differential equations~\eqref{eq:zeq} and~\eqref{eq:logV}, and $S(t)$ with~\eqref{eq:btzS}.
The agreement provides a quantitative instance from the point of view of the ER=EPR conjecture~\cite{Maldacena:2013xja,Harlow:2016vwg}: the boundary entanglement between $A=A_L\cup A_R$ and its complement, computed purely in the CFT, grows exactly as the length of the Einstein--Rosen bridge at $x=0$.
At early times $S(t)-S(0)\simeq\frac{c}{3}\frac{2\pi^2}{\beta^2}t^2$, and at late times $S(t)\simeq \frac{c}{3}\frac{2\pi}{\beta}t=2s\,t$, i.e.\ $\vE=1$.
In the Penrose diagram, the HM surfaces of BTZ are the constant-global-time slices of the AdS$_2$ geometry of the $(t,z)$ plane; the turning point $\zt=\cosh t$ goes to the would-be singularity exponentially fast, so that BTZ HM surfaces probe arbitrarily deep regions of the interior at late times.

\subsubsection{Numerical analysis: charged BTZ backgrounds}\label{sec:cbtz}

As a family of nontrivial interiors we consider the charged BTZ black hole~\cite{Martinez:1999qi}, a solution of
\begin{equation}
S_{\rm bulk}=\frac{1}{16\pi\GN}\int\dd^3x\sqrt{-g}\left(R+2-\frac14F_{\mu\nu}F^{\mu\nu}\right),
\end{equation}
with
\begin{equation}\label{eq:cbtz}
f(z)=1-\frac{z^2}{\zh^2}+\mu^2z^2\log\frac{z}{\zh}\,,\qquad \chi=0\,,\qquad A=\sqrt2\,\mu\log\frac{z}{\zh}\,\dd t\,.
\end{equation}
In AdS$_3$ the gauge potential grows logarithmically at the boundary, and its coefficient $\mu$ is the charge parameter.
The temperature
\begin{equation}
T=\frac{2-\mu^2\zh^2}{4\pi\zh}
\end{equation}
vanishes at $\mu\zh=\sqrt2$, where the outer horizon merges with the inner (Cauchy) horizon $z_->\zh$.
The effective potential $\Veff=\zh^{-2}-z^{-2}-\mu^2\log(z/\zh)$ now has a maximum between the two horizons, and~\eqref{eq:zstar} and~\eqref{eq:vE} give
\begin{equation}\label{eq:cbtzvE}
\zs=\frac{\sqrt2}{\mu}\,,\qquad
\vE^2=1-\frac{\mu^2\zh^2}{2}-\frac{\mu^2\zh^2}{2}\log\frac{2}{\mu^2\zh^2}\,.
\end{equation}
Thus $\vE<1$ for $\mu>0$, and both $\vE$ and the depth $\zs-\zh$ of the interior accessible to HM surfaces shrink to zero in the extremal limit ($\mu\zh=\sqrt2$).

\begin{figure}[t]
\centering
\includegraphics[width=0.92\textwidth]{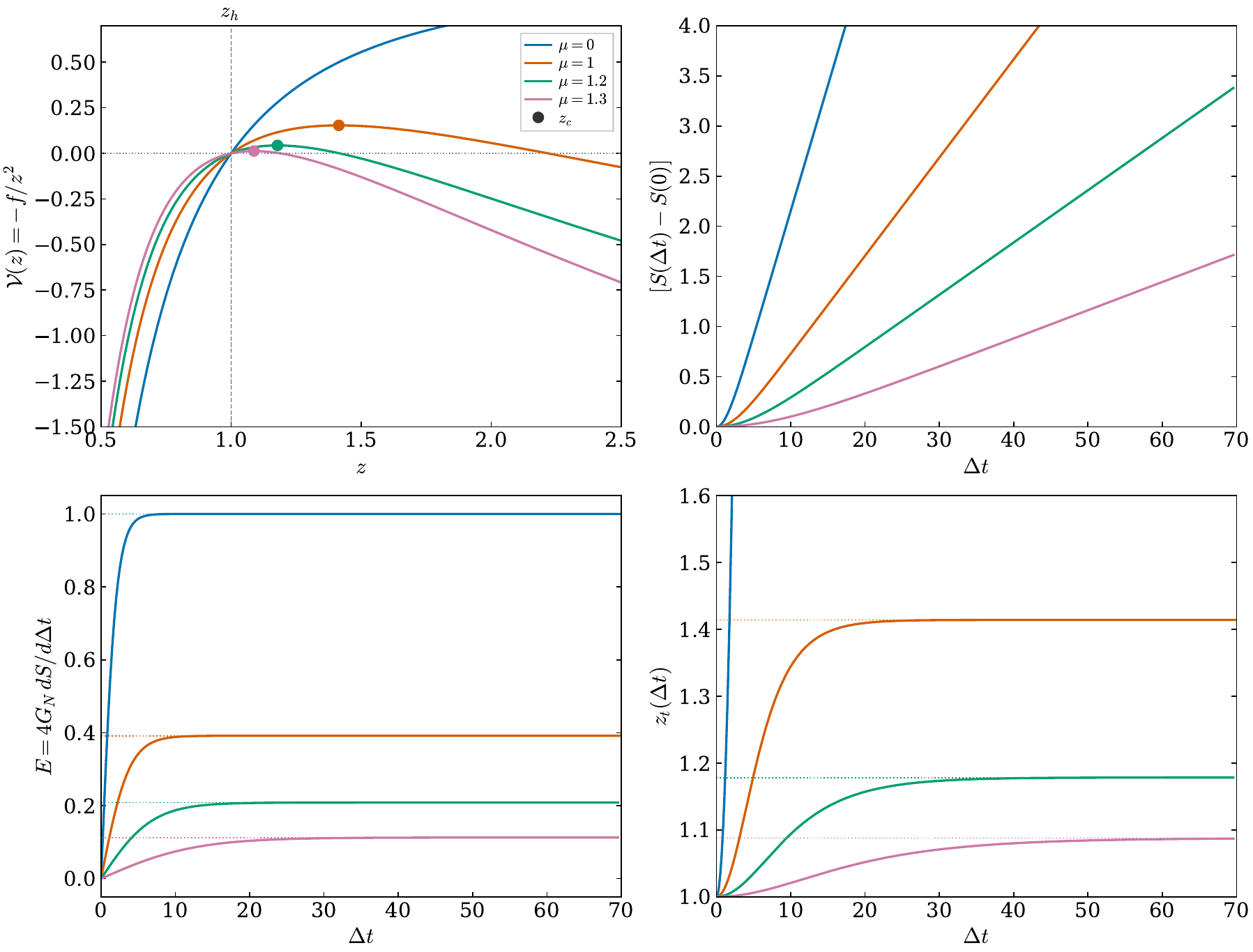}
\caption{Entanglement dynamics of charged BTZ black holes ($\zh=1$) for $\mu=0,1,1.2,1.3$, against the total boundary time $\Delta t=2t$. \textbf{Top left:} effective potential $\Veff(z)=-f/z^2$; the dashed line marks the horizon and dots mark the critical surfaces $\zs$. \textbf{Top right:} entanglement growth $S(\Delta t)-S(0) = S_{\rm reg}(\Delta t)-S_{\rm reg}(0)$ ($\GN=1$). \textbf{Bottom left:} conserved energy $E=4\GN\,\dd S/\dd\Delta t=2\GN\,\dd S/\dd t$, which approaches $\vE$ (dotted lines) at late times. \textbf{Bottom right:} turning point $\zt$, which approaches $\zs$ (dotted lines); for $\mu=0$, $\zt=\cosh(\Delta t/2)$.}
\label{fig:cbtz}
\end{figure}

Figure~\ref{fig:cbtz} summarises the entanglement dynamics.
The effective potential (top left) shows how the barrier height $E_c^2=\vE^2$ and its location $\zs$ decrease with $\mu$.
After an initial quadratic growth, $S(t)$ (top right) becomes linear, with the slope $\dd S/\dd t=2s\vE$ reached when the energy $E(t)$, obtained from the slope through~\eqref{eq:HJ}, saturates at $E_c$ (bottom left).
Simultaneously, the turning point approaches the critical surface (bottom right).
The approach is exponential in time, and the relevant time scale grows as the temperature decreases.
The curves of $S(\Delta t)$ are the input data for the reconstruction in section~\ref{sec:recon}: each value of $\Delta t$ probes the geometry down to $\zt(\Delta t)$, and the late-time regime is controlled by the neighbourhood of $\zs$.

%
\section{Bulk reconstruction from entanglement dynamics}\label{sec:recon}

We now turn to the inverse problem: given the entanglement entropy $S(t)$ of section~\ref{sec:hm}, find the metric functions $f(z)$ and $\chi(z)$.

%
\subsection{Limitations of Abel inversion in bulk reconstruction}\label{sec:abel}

\paragraph{Static entanglement.}
It is instructive to recall first why the static problem can be solved without machine learning.
For an interval of length $\ell$ on one boundary at fixed $t$, the RT geodesic with turning point $z_\ell$ has
\begin{equation}
\ell(z_\ell)=2\int_0^{z_\ell}\dd z\,\frac{z}{\sqrt{f(z)}\sqrt{z_\ell^2-z^2}}\,,\qquad
\frac{\dd S}{\dd\ell}=\frac{1}{4\GN z_\ell}\,,
\end{equation}
where the second relation is the Hamilton--Jacobi relation for static surfaces.
Two features make this problem elementary.
First, the turning point is read off \emph{directly} from the data, $z_\ell=(4\GN\,\dd S/\dd\ell)^{-1}$, independently of the metric.
Second, with $w=z^2$ the first relation becomes $\ell=\int_0^{w_\ell}\dd w\,f^{-1/2}/\sqrt{w_\ell-w}$, an Abel equation for $f^{-1/2}$, which is inverted in closed form:
$f(w)^{-1/2}=\pi^{-1}\partial_w\int_0^w\dd w_\ell\,\ell(w_\ell)/\sqrt{w-w_\ell}$.
This is Bilson's method~\cite{Bilson:2008ab,Bilson:2010ff}.
Since static geodesics never see $g_{tt}$, $S(\ell)$ determines $f(z)$ outside the horizon whatever $\chi$ is, and when $\chi=0$ it determines the full exterior metric~\cite{Ahn:2024lkk}.

\paragraph{Entanglement dynamics.}
For HM surfaces the Hamilton--Jacobi relation~\eqref{eq:HJ} also turns the data into a property of the surface, but the property is the energy $E$, not the turning point: by~\eqref{eq:turning}, $\zt$ is related to $E$ through the unknown metric, even when $\chi=0$.
To see what the data actually determine, we change variables from $z$ to the Killing norm $\Veff$, which is monotonic between the boundary and $\zs$.
We define the \emph{local surface gravity}
\begin{equation}\label{eq:kappa}
\kappa\equiv\frac12\,z^2e^{\chi/2}\,\frac{\dd\Veff}{\dd z}\,,\qquad\text{equivalently}\qquad
\kappa^2=-\frac{\nabla_\mu\Veff\,\nabla^\mu\Veff}{4\Veff}\,,
\end{equation}
which equals the surface gravity $2\pi T$ at the horizon and vanishes at $\zs$.
The measure in~\eqref{eq:AE} then becomes $e^{-\chi/2}\dd z/z^2=\dd\Veff/(2\kappa)$, and~\eqref{eq:AE}, \eqref{eq:tE} become
\begin{equation}\label{eq:abelform}
4\GN S(E)=\int_{-\infty}^{E^2}\frac{\dd\Veff}{\kappa(\Veff)\sqrt{E^2-\Veff}}\,,\qquad
t(E)=\frac{E}{2}\;\PV\!\int_{-\infty}^{E^2}\frac{\dd\Veff}{\Veff\,\kappa(\Veff)\sqrt{E^2-\Veff}}\,,
\end{equation}
where the first integral is regularised by the cutoff at $\Veff(\epsilon)\simeq-1/\epsilon^2$.

The integral representation in~\eqref{eq:abelform} shows that both $S(E)$ and $t(E)$ depend on the bulk metric only through the single function $\kappa(\Veff)$. Consequently, two geometries with the same $\kappa(\Veff)$ for $\Veff\in(-\infty,E_c^2]$ yield identical half-space entanglement dynamics,
\begin{equation}
  \kappa_1(\Veff)=\kappa_2(\Veff) \quad\Longrightarrow\quad S_1(t)=S_2(t)\,.
  \label{eq:kappaequiv}
\end{equation}
The inverse problem therefore splits into two steps,
\begin{equation}
  S(t)\;\xrightarrow{\ \text{(i)}\ }\;\kappa(\Veff)\;\xrightarrow{\ \text{(ii)}\ }\;(f,\chi)\,,
  \label{eq:two-steps}
\end{equation}
which behave very differently: step~(ii) is a question of uniqueness, step~(i) one of
invertibility.

\paragraph{Step (ii): from $\kappa(\Veff)$ to the metric.}
For $\chi=0$, the definition \eqref{eq:kappa} gives $\kappa=f/z-f'/2$ and
$\dd(1/z)=-\dd\Veff/2\kappa$. If $\kappa(\Veff)$ is known, integrating the latter from the
horizon ($\Veff=0$, $z=z_h$) gives $z(\Veff)$, which is monotonic for $z<z_c$ and can be
inverted to $\Veff(z)$; the definition of $\Veff$ then returns $f=-z^2\Veff(z)$. Step~(ii)
is thus unique and explicit once $z_h$ is fixed. For example, BTZ
has $\Veff=1-1/z^2$ and $\kappa=1/z=\sqrt{1-\Veff}$, from which this procedure returns
$f=1-z^2$; for charged BTZ, $\kappa=1/z-\mu^2z/2$, which equals $\kappa_h$ at the horizon
and vanishes at $z_c=\sqrt2/\mu$. When $\chi\neq0$, by contrast, a single function
$\kappa(\Veff)$ cannot fix both $f$ and $\chi$, and step~(ii) is not unique.

\paragraph{Step (i): from $S(t)$ to $\kappa(\Veff)$.}
The first relation in \eqref{eq:abelform} is an Abel transform of $1/\kappa$. If $S$ were known for all $\varepsilon\equiv E^2\in(-\infty,E_c^2)$, it could be inverted
in closed form, as in the static case,
\begin{equation}
  \frac{1}{\kappa(\Veff)}=\frac{1}{\pi}\frac{\dd}{\dd\Veff}\int_{-\infty}^{\Veff}
  \dd\varepsilon\,\frac{4G_N S(\varepsilon)}{\sqrt{\Veff-\varepsilon}}\,.
  \label{eq:abel-inverse}
\end{equation}
Real-time data, however, only provide $\varepsilon\geq0$. Negative $\varepsilon$
corresponds to imaginary $E$, i.e.\ to boundary times continued to imaginary values
$t=i\tau$, for which the two endpoints sit on the Euclidean thermal circle and the
geodesic stays outside the horizon. Splitting the integral in \eqref{eq:abelform} at $\Veff=0$,
\begin{equation}
  4G_N S(\varepsilon)
  =\underbrace{\int_{-\infty}^{0}\frac{\dd\Veff}{\kappa\sqrt{\varepsilon-\Veff}}}_{\text{exterior}}
  +\underbrace{\int_{0}^{\varepsilon}\frac{\dd\Veff}{\kappa\sqrt{\varepsilon-\Veff}}}_{\text{interior}}\,,
  \label{eq:abelsplit}
\end{equation}
we see that the interior enters through an Abel kernel, whose singularity at
$\Veff=\varepsilon$ ties each value of $\varepsilon$ to the geometry near the turning
point. The exterior instead enters through a generalized Stieltjes transform of order
$1/2$, $\int_0^\infty\dd s\,\kappa(-s)^{-1}(\varepsilon+s)^{-1/2}$, which is a smooth
function of $\varepsilon>0$: real-time surfaces cross the exterior without turning there.
The exterior geometry is thus encoded in real-time data only through the analytic
continuation of $S$ to $\varepsilon<0$, i.e.\ to imaginary time, a problem that is
exponentially ill-conditioned.
Hence, in contrast to the static case, where
Bilson's Abel inversion yields $f$ in closed form whenever $\chi=0$, step~(i) has no
closed-form solution, \emph{even when $\chi=0$}.

In essence, eq.~\eqref{eq:abelform} shows that $S(t)$ contains one function's worth of information $\kappa(\Veff)$, so that the two functions $(f,\chi)$ cannot both be determined.
All these observations set the stage for this section: we need a global, nonlinear and physics-informed fit to determine $f$ when $\chi=0$, and we should expect a ``degeneracy" of metric when $\chi$ is free (see section~\ref{sec:learnf}).

%
\subsection{Deep learning black hole spacetime}\label{sec:dl}

\subsubsection{Neural ordinary differential equations: a method}\label{sec:node}

\paragraph{Neural ODEs and the adjoint method.}
A residual network updates its hidden state as $X_{n+1}=X_n+h\,F(X_n;\theta_n)$.
In the limit of many layers and small $h$ this becomes an ODE in a continuous depth variable $\Lambda$~\cite{E:2017dyn,Haber:2017stable,Lu:2017beyond,Chen:2018neural},
\begin{equation}\label{eq:node}
\frac{\dd X}{\dd\Lambda}=F\big(X(\Lambda),\Lambda;\theta\big)\,,\qquad X(\Lambda_0)=X_0(\theta)\,,
\end{equation}
whose forward pass is the solution of an initial-value problem and whose output $X(\Lambda_1)$ enters a loss $\Loss$.
The gradient $\dd\Loss/\dd\theta$ is obtained from the \emph{backward dynamics} of the adjoint state $a(\Lambda)\equiv\partial\Loss/\partial X(\Lambda)$~\cite{Pontryagin:1962,Chen:2018neural}.
Introducing a Lagrange multiplier for~\eqref{eq:node} and integrating by parts, one finds that $a$, treated as a row vector, obeys
\begin{equation}\label{eq:adjoint}
\frac{\dd a}{\dd\Lambda}=-\,a\,\frac{\partial F}{\partial X}\,,\qquad a(\Lambda_1)=\frac{\partial\Loss}{\partial X(\Lambda_1)}\,,
\end{equation}
which is integrated backwards from $\Lambda_1$ to $\Lambda_0$, together with $X(\Lambda)$ itself (reconstructed by integrating~\eqref{eq:node} backwards, so that no intermediate states need to be stored), while the gradient accumulates as
\begin{equation}\label{eq:adjointgrad}
\frac{\dd\Loss}{\dd\theta}=\int_{\Lambda_0}^{\Lambda_1}\dd\Lambda\;a\,\frac{\partial F}{\partial\theta}+a(\Lambda_0)\frac{\partial X_0}{\partial\theta}\,.
\end{equation}
Equations~\eqref{eq:adjoint}--\eqref{eq:adjointgrad} are the continuum version of backpropagation: the backward pass is itself an ODE, with memory cost independent of depth.
The alternative, which we use, is to differentiate the discretised forward pass directly, by reverse-mode automatic differentiation through the steps of the ODE solver; it gives the exact gradient of the numerical solution, at the cost of storing the intermediate states.

\paragraph{Holographic neural ODE.}
In our case the forward pass is the HM surface of section~\ref{sec:differential}: \textit{the depth variable $\Lambda$ is the proper length $\lambda$}, the state is $X=(z,z',\log|V|)$, and the vector field $F$, with the model metric functions $f_\theta$, $\chi_\theta$ and the surface gravity $\kappa_{h,\theta}$ specified in~\eqref{eq:ansatz} below, is
\begin{equation}\label{eq:F}
F=\left(z'\,,\;\frac12\,\partial_z\Big[z^2f_\theta+E^2z^4e^{\chi_\theta}\Big]\,,\;\frac{\kappa_{h,\theta}\,z^2}{Ez^2-e^{-\chi_\theta/2}z'}\right),
\end{equation}
where the second and third components are~\eqref{eq:zeq} and~\eqref{eq:logV}.
The unknown metric enters only through the vector field~\eqref{eq:F}, and is represented by two deep neural networks $D_f$ and $D_\chi$ through
\begin{equation}\label{eq:ansatz}
f_\theta(z)=(1-z)\big(1+z\,D_f(z)\big)\,,\qquad \chi_\theta(z)=z\,D_\chi(z)\,.
\end{equation}
The ansatz builds in only the boundary normalisation $f(0)=1$, $\chi(0)=0$ and the horizon at $\zh=1$.
The temperature is learned through $f_\theta'(\zh)=-1-D_f(1)$, so that $\kappa_{h,\theta}=\tfrac12\big(1+D_f(1)\big)e^{-\chi_\theta(1)/2}$.
Each network is fully connected with four hidden layers of $10$ neurons and softplus activations, and the trainable parameters $\theta$ are the weights and biases of the two networks; when $\chi$ is not learned, $\chi_\theta=0$.
A smooth activation is required because the vector field~\eqref{eq:F} involves $f'$ and its Jacobian $\partial F/\partial X$ involves $f''$.

\paragraph{Loss function.}
The network parameters $\theta$ are trained by minimising the data loss
\begin{equation}\label{eq:loss}
\Loss(\theta)=\frac{1}{N}\sum_{i=1}^{N}\log\cosh\big[S_{\rm reg}^{\theta}(t_i)-S_{\rm reg}(t_i)\big]\,,
\end{equation}
which is the sole training objective; for small residuals, $\log\cosh x\simeq x^2/2$. Here $S_{\rm reg}(t_i)$ are the input entanglement data at $N$ boundary times $t_i$, and $S_{\rm reg}^{\theta}(t_i)$ is the prediction of the current metric $\{f_\theta,\chi_\theta\}$ in~\eqref{eq:ansatz}; we set $\GN=1$. The prediction is obtained in three steps.
\begin{enumerate}
\item \emph{Energies.} HM surfaces are labelled by their energy $E$ rather than by $t$: $E=0$ gives $t=0$, and $t\to\infty$ as $E\to E_c$. For the current metric we compute $E_{c,\theta}^2=\Veff(\zs)$, where $\zs$ solves~\eqref{eq:zstar} and $\Veff$ is defined in~\eqref{eq:tprime}, and take 80 energies $E_j$ with $E_j^2\in E_{c,\theta}^2\times[10^{-3},1-10^{-8}]$.\footnote{The normalized squared energies $u_j\equiv E_j^2/E_{c,\theta}^2$ are uniformly spaced on a complementary log-log scale over $u_j\in[10^{-3},1-10^{-8}]$.} Since $E_{c,\theta}$ depends on $f_\theta$ and $\chi_\theta$, the energies are updated at every training step.
\item \emph{Surfaces.} For each $E_j$ we solve the HM surface equations~\eqref{eq:zeq}--\eqref{eq:logV} from the horizon as the initial condition~\eqref{eq:z1E}, and read off the boundary time $t_\theta(E_j)$ and the regularised entropy $S_{\rm reg}^{\theta}(E_j)$ from~\eqref{eq:tV}. The integration is stopped near the boundary, at $z=\epsilon \simeq e^{-10}$.
\item \emph{Interpolation.} The times $t_\theta(E_j)$ do not coincide with the data times $t_i$. We therefore interpolate the points $\big(t_\theta(E_j),S_{\rm reg}^{\theta}(E_j)\big)$ using their exact slopes $\dd S/\dd t=E_j/2\GN$ from~\eqref{eq:HJ}, and evaluate the result at $t_i$. We set $i=1,\cdots, 200$.
\end{enumerate}

\begin{figure}[]
\centering
\includegraphics[width=\textwidth]{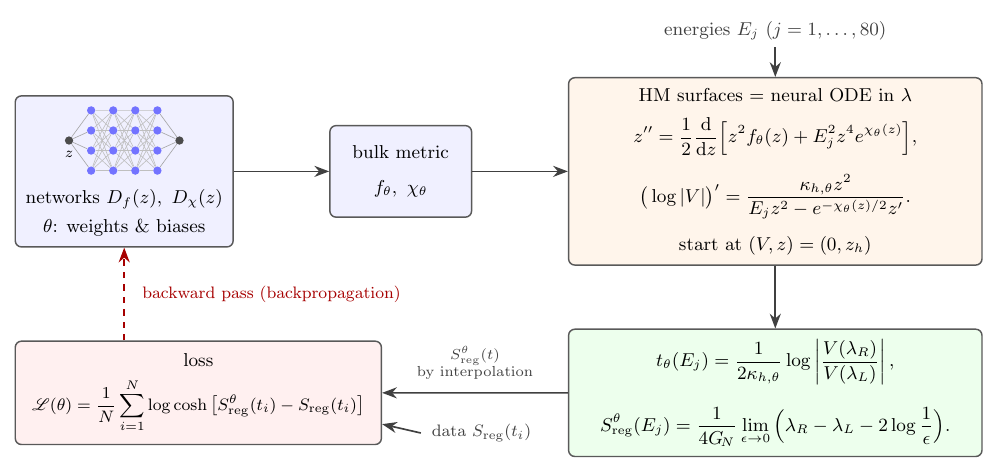}
\caption{Schematic of the neural ODE framework for bulk reconstruction from HM entanglement dynamics. The networks $D_f$ and $D_\chi$ parametrise the bulk metric~\eqref{eq:ansatz}, which enters the HM surface equations~\eqref{eq:zeq}--\eqref{eq:logV}. For each energy $E_j$, the equations are integrated from the horizon to obtain $t_\theta(E_j)$ and $S_{\rm reg}^{\theta}(E_j)$ via~\eqref{eq:tV}, and interpolation yields $S_{\rm reg}^{\theta}(t)$. The predicted entropy is compared with the data $S_{\rm reg}(t_i)$ through the loss~\eqref{eq:loss}, and the network parameters are updated by backpropagation.}
\label{fig:schematic}
\end{figure}

\paragraph{Data.}
The data $S_{\rm reg}(t_i)$ are given at $N$ times spread uniformly over $0<t_i\le t_{\max}$. We choose $t_{\max}$ as the time at which the turning point $\zt$ has covered $99\%$ of the way from $\zh$ to $\zs$, so that the data include the late-time regime in which $\zt\to\zs$ and $S$ grows linearly. The network architecture and training procedure are illustrated in figure~\ref{fig:schematic}.

\paragraph{Training setup.}
Each surface is launched at $\lambda=\pm\delta$, $\delta=10^{-8}$, with the initial data of section~\ref{sec:differential}, $V_{(1)}=1$ and $z_{(1)}=e^{\chi_\theta(1)/2}E$, and its two sides, $\lambda>0$ and $\lambda<0$, are integrated separately towards the right and left boundaries. Both sides of all 80 surfaces are integrated together as one batch with the adaptive Dormand--Prince scheme~\cite{Dormand:1980} of \texttt{torchode}~\cite{Lienen:2022aa}. 

The networks are initialised with the default random initialisation of PyTorch~\cite{Paszke:2019pytorch}. Runs with different random seeds start from different metrics; for each background we report a single run. We first train with Adam~\cite{Kingma:2014adam} with learning rate $10^{-3}$, taking one step per evaluation of the full loss, and switch to L-BFGS once $\Loss<10^{-4}$, or, if this value is not reached (as for the superluminal data in the next section), once the loss stops decreasing. L-BFGS is continued until the loss stops decreasing, and we report the model with the lowest loss. The absolute and relative solver tolerances are $(10^{-9},10^{-6})$ during the Adam stage, and $(10^{-11},10^{-8})$ during the L-BFGS stage and for all reported results.

\begin{figure}[t]
\centering
\includegraphics[width=\textwidth]{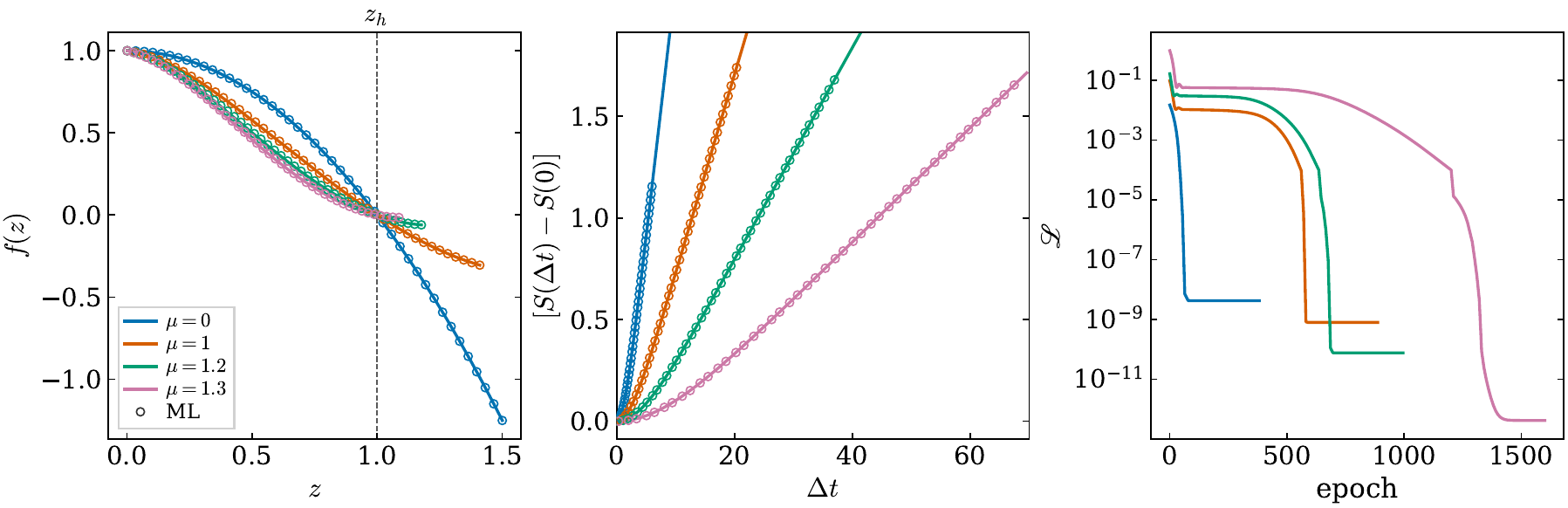}
\caption{Reconstruction of charged BTZ black holes from the dynamical entanglement entropy data when $\chi=0$ ($\zh=1$) for $\mu=0,1,1.2,1.3$ (colors). \textbf{Left:} learned $f_\theta(z)$ (circles) compared with~\eqref{eq:cbtz} (lines) on the probed region $0<z<\zt(t_{\max})$, shown up to $z=1.5$ for $\mu=0$; the dashed line marks the horizon $z=\zh$. \textbf{Middle:} $S(\Delta t)-S(0) = S_{\rm reg}(\Delta t)-S_{\rm reg}(0)$ ($\GN=1$) against $\Delta t$, data (lines) and the model $S_{\rm reg}^{\theta}(t_i)$ at the data times (circles). \textbf{Right:} training loss~\eqref{eq:loss} against the number of epochs (Adam epochs followed by L-BFGS iterations).}
\label{fig:learnf}
\end{figure}

%
\subsubsection{Learning black hole spacetime from entanglement dynamics}\label{sec:learnf}

\paragraph{Reconstruction of the charged BTZ black hole.}
We first set $\chi=0$ and learn the blackening factor $f$ alone from the charged BTZ data of section~\ref{sec:cbtz}, for $\mu=0,1,1.2,1.3$. The results are summarised in figure~\ref{fig:learnf}. Starting from random initial metrics, the loss decreases by many orders of magnitude in all cases, and the L-BFGS stage converges to $\Loss\sim10^{-13}$--$10^{-9}$. The learned $f_\theta$ agrees with the true blackening factor~\eqref{eq:cbtz} over the whole region probed by the HM surfaces, $0<z<\zt(t_{\max}) \simeq \zs$,\footnote{For BTZ ($\mu=0$), $\zs=\infty$, and the probed region is $0<z<\zt(t_{\max})\simeq10.1$.} including the interior behind the horizon. For $\chi=0$, the entanglement dynamics alone therefore determines the dual black hole geometry in the probed region.

This outcome is not guaranteed by the analysis of section~\ref{sec:abel}. There we found that for $\chi=0$ the inverse problem has no closed-form solution, and that it is unique only for analytic metrics, through a non-local and ill-conditioned dependence on the exterior. The network nevertheless converges to the correct metric from random initial conditions. Since the softplus activation is real-analytic, every $f_\theta$ is an analytic function of $z$, so the network searches within the class of metrics for which the inverse problem is unique; this excludes the smooth but non-analytic deformations that leave $S(t)$ unchanged.

\begin{figure}[t]
\centering
\includegraphics[width=0.92\textwidth]{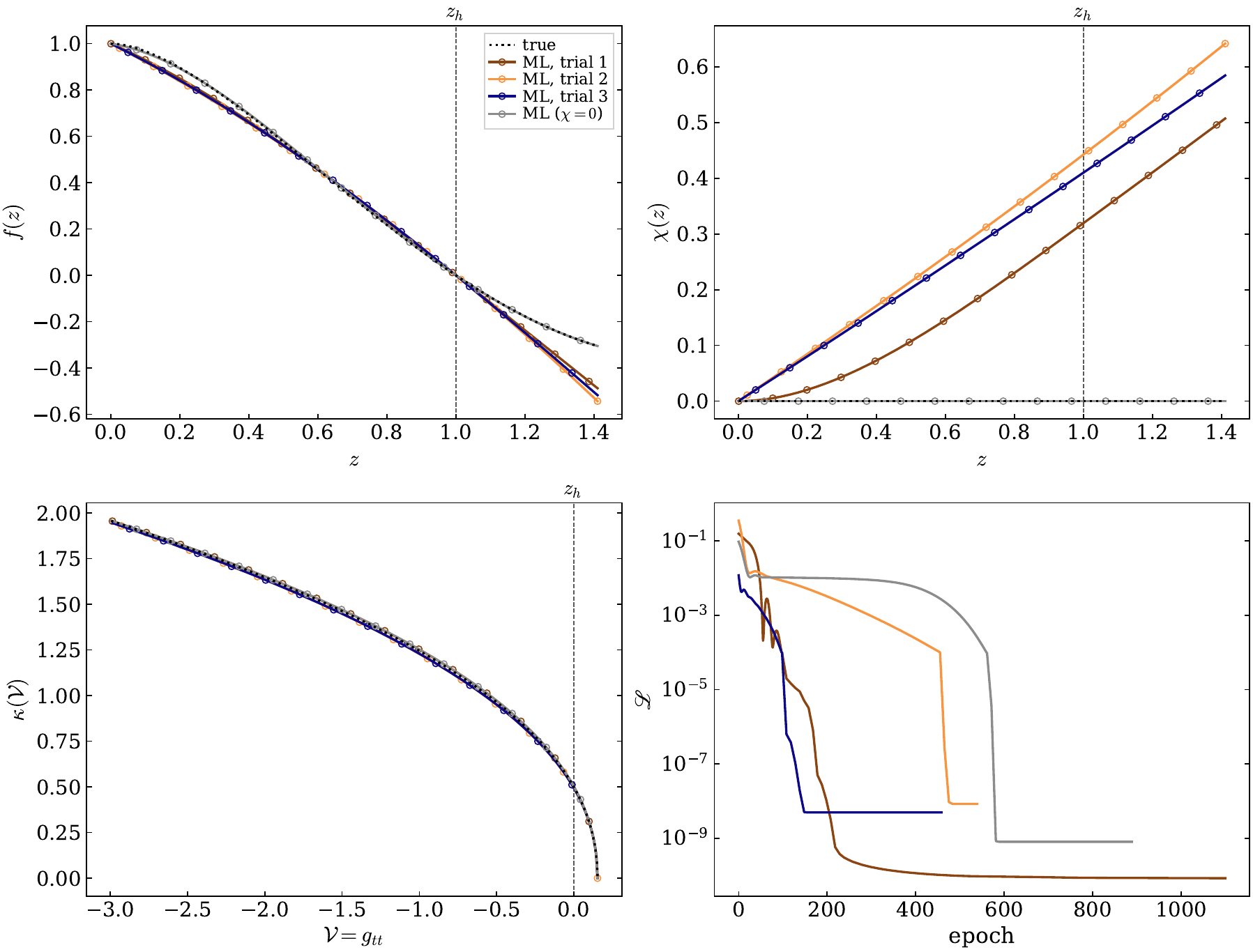}
\caption{Degeneracy when $f$ and $\chi$ are learned together from the $\mu=1$ entanglement entropy data. Three trainings are shown (lines with circles): trial~1 with $\chi_\theta=z^2D_\chi$, and trials~2 and~3 with $\chi_\theta=zD_\chi$ and different random seeds. For comparison, the $\chi=0$ reconstruction is shown in grey and the true geometry~\eqref{eq:cbtz} as dotted lines. \textbf{Top:} learned $f_\theta(z)$ and $\chi_\theta(z)$ for $0<z<\zt(t_{\max})\simeq\zs$, the region probed by the data in the true geometry. \textbf{Bottom left:} local surface gravity $\kappa$~\eqref{eq:kappa} against the Killing norm $\Veff=g_{tt}$. Within the training accuracy, all learned geometries share the true $\kappa(\Veff)$, the only combination of the metric on which $S(t)$ depends. \textbf{Bottom right:} training loss~\eqref{eq:loss} against the number of epochs (Adam epochs followed by L-BFGS iterations). In the top and bottom-left panels, the dashed lines mark the horizon, $z=\zh$ and $\Veff=0$.}
\label{fig:degeneracy}
\end{figure}

\paragraph{Degeneracy in reconstructed spacetimes.}
We now learn both metric functions, $f_\theta$ and $\chi_\theta$, from the same dynamical entanglement entropy data: as the representative example, we use the data when $\mu=1$. Figure~\ref{fig:degeneracy} shows three such trainings, together with the $\chi=0$ reconstruction above from the same data. In trial~1 we use $\chi_\theta=z^2D_\chi(z)$, which imposes $\chi_\theta'(0)=0$ as for the true geometry; this tests whether matching the near-boundary behaviour of the true metric is enough to remove any ambiguity. Trials~2 and~3 use the ansatz~\eqref{eq:ansatz}, in which $\chi_\theta'(0)$ is not fixed, and differ only in the random initialisation of the networks.

All three trainings reproduce the data, with final losses $\Loss\sim10^{-11}$--$10^{-9}$, at the same level as the $\chi=0$ reconstruction ($\Loss\sim10^{-9}$). Yet the learned geometries differ from the true one and from each other. In trial~1 the learned metric stays close to the true one near the boundary but deviates strongly near and inside the horizon, and trials~2 and~3, which differ only in their initialisation, converge to different geometries. Hence $S(t)$ alone does not determine the metric once $\chi$ is free.

\paragraph{The degeneracy is a diffeomorphism of the $(t,z)$ plane.}
Equation~\eqref{eq:abelform} explains these observations. Since $S(t)$ only knows $\kappa(\Veff)$, any two metrics with the same $\kappa(\Veff)$ are indistinguishable, and figure~\ref{fig:degeneracy}(bottom left) confirms that the learned metrics all share the exact $\kappa(\Veff)$ \eqref{eq:kappa}. Geometrically, the HM surface of half-spaces lies at $x=0$, so its length is computed entirely within the two-dimensional metric of the $(t,z)$ plane,
\begin{equation}
\dd s_2^2=\frac{1}{z^2}\left(-f e^{-\chi}\dd t^2+\frac{\dd z^2}{f}\right),
\end{equation}
and never feels the transverse metric $g_{xx}=1/z^2$.
A static two-dimensional metric with a Killing time normalised at the boundary contains a single function's worth of diffeomorphism-invariant information, which is precisely $\kappa(\Veff)$: $\Veff=g_{tt}$ and $-\nabla_\mu\Veff\nabla^\mu\Veff/4\Veff=\kappa^2$ are scalars of $\dd s_2^2$.
The ansatz~\eqref{eq:metric}, however, uses $g_{xx}$ to \emph{define} the radial coordinate.
A reparametrisation $z\to Z(z)$ of the $(t,z)$ plane therefore maps one solution of the inverse problem to another, with a different $g_{xx}$ relative to $g_{tt}$ and $g_{zz}$.
Explicitly, for any smooth monotonic $Z(z)$ one finds
\begin{equation}\label{eq:degenerate}
\tilde f\big(Z(z)\big)=\frac{z^2Z'(z)^2}{Z(z)^2}\,f(z)\,,\qquad
\tilde\chi\big(Z(z)\big)=\chi(z)+2\log\frac{z^2Z'(z)}{Z(z)^2}
\end{equation}
gives $\dd\tilde s_2^2(Z)=\dd s_2^2(z)$, hence exactly the same $S(t)$ with $Z(0)=0$, $Z'(0)=1$ and $Z(\zh)=\zh$, while $\dd x^2/Z^2\neq\dd x^2/z^2$, so that the three-dimensional geometries are \textit{physically different}. In other words, the flat directions of the loss are identified as a bulk symmetry of the probe, rather than as a numerical artifact.

\paragraph{Causality and superluminal entanglement.}
In the reconstructions so far the data came from ``healthy" geometries. We now ask what the network does when it is fed \emph{unphysical} entanglement dynamics, namely an entropy that grows faster than allowed by causality. We set $\chi=0$ and learn $f$ only for simplicity.

The relevant bound is known. In holographic theories whose bulk obeys the null energy condition (NEC), the entanglement velocity cannot exceed its value for the neutral black brane~\cite{Mezei:2016wfz,Mezei:2016zxg}, which in three bulk dimensions is the BTZ value $v_E=1$. The same bound, $v_E\le1$, holds in any relativistic quantum field theory~\cite{Hartman:2015apr}, and in holography the NEC is also what guarantees boundary causality~\cite{Gao:2000ga}. For $\chi=0$ the NEC reads \begin{equation}\label{eq:nec} zf''(z)-f'(z)\geq0 \end{equation} both outside and inside the horizon; charged BTZ satisfies it, with $zf''-f'=2\mu^2z$, and indeed has $v_E<1$. Entanglement data with $v_E>1$ therefore cannot come from a geometry with $\chi=0$ that obeys the NEC. Our network, however, does not impose~\eqref{eq:nec}, so the bound is not built into the reconstruction.

We construct artificial data by keeping everything about the BTZ entropy fixed except its late-time slope,
\begin{equation}\label{eq:Sv}
  4\GN S_v(t)=2\log2+2v^2\log\cosh\frac{t}{v}\,,\qquad
  E_v(t)=v\tanh\frac{t}{v}\qquad(\zh=1)\,.
\end{equation}
For every $v$, $S_v(0)$ and the early growth $4\GN[S_v(t)-S_v(0)]\simeq t^2$ coincide with those of BTZ, while at late times $\dd S_v/\dd t\to2sv$, i.e.\ $v_E=v$. The case $v=1$ is exactly the BTZ data; data with $v<1$ respect the bound, and data with $v>1$ violate it. We train with the setup of section~\ref{sec:node} for $v=0.6$, $0.8$, $1$, $1.2$ and $1.4$, on $N=200$ equally spaced points up to $t_{\max}=2.9\,v$, chosen so that every data set reaches the same fraction of its late-time slope, $E_v(t_{\max})=0.994\,v$. 

\begin{figure}[t]
\centering
\includegraphics[width=0.92\textwidth]{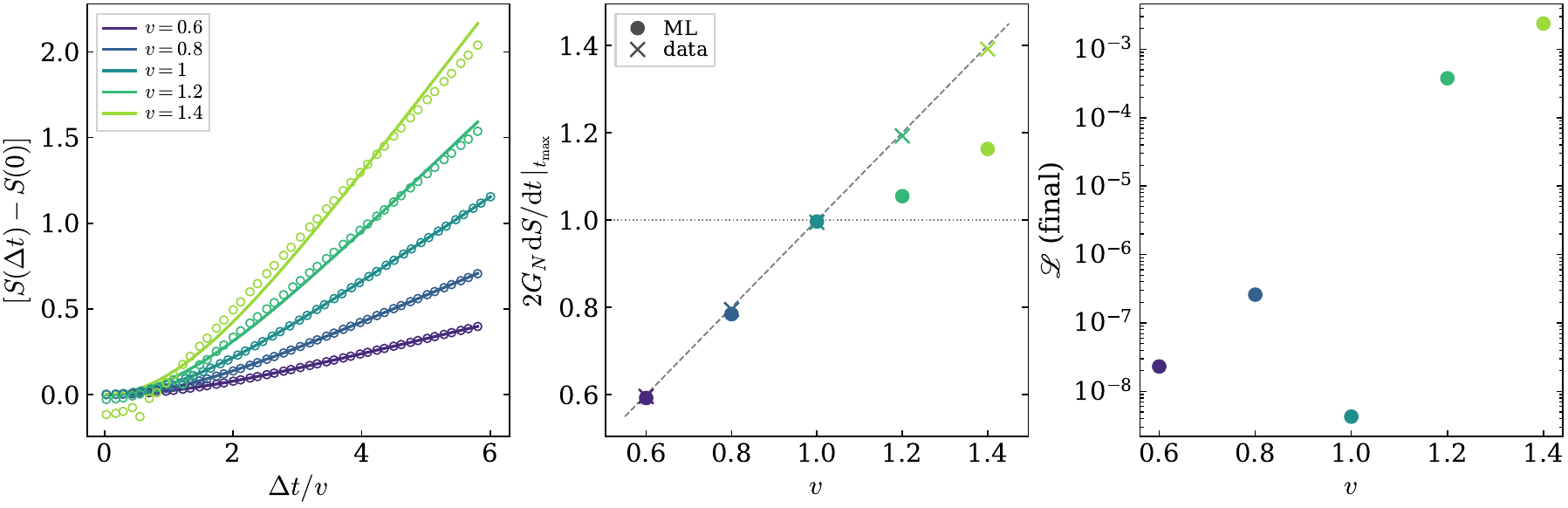}
\caption{Learning from the artificial entanglement data~\eqref{eq:Sv} with $\chi=0$, for $v=0.6,0.8,1,1.2,1.4$; $v=1$ is the BTZ run above. \textbf{Left:} $S(\Delta t)-S(0) = S_{\rm reg}(\Delta t)-S_{\rm reg}(0)$ ($\GN=1$) of the data (lines) and of the learned model at the data times (circles), against $\Delta t/v$. \textbf{Middle:} slope $2\GN\,\dd S/\dd t$ at the end of the data window, $t=t_{\max}$, of the learned model (filled circles) and of the data (crosses); the dashed line is $v$, and the dotted line marks the causal bound $v_E=1$. \textbf{Right:} final value of the loss~\eqref{eq:loss}.}
\label{fig:superluminal}
\end{figure}

The results, shown in figure~\ref{fig:superluminal}, separate sharply at the causal bound. For $v\le1$ the network reproduces the data, with final losses between $4\times10^{-9}$ and $3\times10^{-7}$, and the slope of the learned entropy at the end of the window agrees with that of the data to within $1.4\%$. For $v>1$ the training struggles: the loss stops decreasing at $\Loss\simeq4\times10^{-4}$ and $2\times10^{-3}$ for $v=1.2$ and $1.4$, at least three orders of magnitude above the subluminal cases, and the entropy deviates visibly from the data, most strongly for $v=1.4$. The learned slope at $t_{\max}$, $1.055$ and $1.163$, stays below that of the data, $1.193$ and $1.392$.

The training thus recovers the causal bound from the data alone: entanglement dynamics with $v_E\le1$ is reproduced, while dynamics with $v_E>1$ is not, even though the network is free to violate the NEC. A residual loss that cannot be removed therefore flags unphysical input. Since the network represents only a restricted class of metrics, we report our result as a property of the present reconstruction, rather than as a general no-go statement.

%
\section{Conclusions}\label{sec:conclusion}
In this work we asked how much of a black hole spacetime is encoded in the time-dependent entanglement entropy of the thermofield double (TFD) state, and how much of it can be recovered in practice. We addressed this question in the simplest setting in which it is sharp: a pair of half-lines, one on each boundary of a three-dimensional eternal black hole, whose HRT surface is the Hartman--Maldacena (HM) surface threading the Einstein--Rosen bridge. Because this surface passes behind the event horizon, the entropy $S(t)$ is sensitive to the black hole interior, in contrast to the static entanglement entropy used in previous reconstructions~\cite{Ahn:2024lkk}, which probes only the exterior. Our answer has three parts. The entropy $S(t)$ depends on the geometry only through a single function; when the metric contains one unknown function, this is enough to reconstruct it, interior included; when it contains two, a degeneracy remains, which we identified as a symmetry of the probe rather than a limitation of the method.

The technical basis of our analysis is the differential formulation of HM surfaces in ingoing Kruskal-like coordinates. In these coordinates the horizon is a regular point of the evolution, so the surfaces are followed across it without the principal-value prescription of the integral formalism and without locating their turning points. The Hamilton--Jacobi relation \eqref{eq:HJ} identifies the conserved energy of each surface, its only free parameter at the horizon, with the slope of $S(t)$. Computing an HM surface therefore becomes an initial-value problem labelled directly by boundary data, which is precisely the structure of a neural ordinary differential equation (neural ODE), with the proper length along the surface playing the role of a continuous network depth. We validated this formulation against the analytic BTZ solution, which reproduces the universal CFT$_2$ result, and against the numerical charged BTZ black holes solutions.

Rewriting the area and the boundary time of HM surfaces in terms of the timelike Killing norm $\mathcal{V}=g_{tt}$ shows that the half-line entanglement dynamics depends on the bulk metric only through the local surface gravity $\kappa(\mathcal{V})$, eq. \eqref{eq:abelform}. This single-function dependence organises the inverse problem into two steps, $S(t)\to\kappa(\mathcal{V})$ and $\kappa(\mathcal{V})\to(f,\chi)$, which raise questions of invertibility and of uniqueness, respectively. The first step is harder than its static counterpart. For static RT surfaces, the turning point is read off directly from the data, and the Abel inversion yields the exterior blackening factor in closed form~\cite{Bilson:2008ab,Bilson:2010ff,Ahn:2024lkk}. For HM surfaces, the data fix the energy $E$ rather than the turning point, and real-time evolution gives access only to $E^2\ge 0$. The interior then enters through an Abel kernel, whereas the exterior enters through a smooth transform that can be resolved only by analytic continuation to imaginary time, an exponentially ill-conditioned problem. Consequently, no closed-form inversion is available, even when the metric contains a single unknown function.

For $\chi=0$, the neural ODE nevertheless reconstructs the blackening factor $f(z)$ from $S(t)$ alone, starting from random initial metrics, for both neutral and charged BTZ black holes. The agreement holds over the entire region probed by the HM surfaces, from the asymptotic boundary through the horizon to the deepest point reached within the data window, which for charged black holes approaches the critical surface inside the event horizon. The Hawking temperature and the location of the critical surface, neither of which is read off directly from the data or used as a training target, follow from the learned metric and agree closely with their exact values. We attribute the success of the reconstruction, despite the ill-conditioning described above, to the real-analytic parametrisation of the metric: since the softplus activation is analytic, so is every learned $f$, and the network searches within a class of metrics for which the inverse problem is unique, excluding the smooth but non-analytic deformations that leave $S(t)$ unchanged. In this sense the architecture itself supplies the regularity assumption under which the inverse problem becomes well posed.

When $f$ and $\chi$ are learned together, the data no longer select a unique geometry. Independent trainings reproduce $S(t)$ as accurately as in the case $\chi=0$, yet converge to metrics that differ from the true one and from each other, even when the near-boundary behavior $\chi'(0)=0$ of the true metric is imposed. This degeneracy is not a numerical artifact but a symmetry of the probe. The half-line HM surface lies at $x=0$, so its area is computed entirely within the static two-dimensional $(t,z)$ metric, whose diffeomorphism-invariant content, once the Killing time is normalized at the boundary, is the single function $\kappa(\mathcal{V})$. Within training accuracy, all learned metrics share the function $\kappa(\mathcal{V})$ of the true geometry, so the flat directions of the loss correspond to this family. With the horizon held fixed, as in our ansatz, the Hawking temperature, the thermal entropy density and the entanglement velocity are common to all members of the family; thermodynamic data alone therefore cannot lift the degeneracy. As shown in appendix~\ref{app:gen}, both the single-function dependence and the associated family of degenerate geometries persist for planar black holes in any number of dimensions.

Finally, we examined how the reconstruction responds to entanglement dynamics that no healthy geometry can produce. In holographic theories whose bulk satisfies the null energy condition (NEC), the rate of entanglement growth is bounded as $\mathrm{d}S/\mathrm{d}t\le 2s$, i.e. $v_E\le 1$~\cite{Mezei:2016zxg,Mezei:2016wfz}, a bound that also holds in any relativistic quantum field theory~\cite{Hartman:2015apr}. Our network does not impose the NEC. We trained it on phenomenological data \eqref{eq:Sv} that share the initial value and the early-time quadratic growth of the BTZ entropy but approach a prescribed late-time slope, $v_E=v$. The outcome separates sharply at the causal bound. Data with $v\le 1$ are reproduced with small losses and with the correct late-time slope, whereas for $v>1$ the loss stays orders of magnitude above the subluminal cases. The training thus recovers the causal bound from the data alone, and an irreducible loss signals entanglement dynamics that the geometries representable by the network cannot produce. Since this class is restricted by the metric ansatz and the architecture, we regard this as a property of the present reconstruction rather than as a general no-go statement.

Several other extensions suggest themselves. Other proposed probes of the interior, such as holographic quantum complexities including Krylov complexity~\cite{Nandy:2024evd,Rabinovici:2025otw,Jeong:2026gdc}, may carry information complementary to entanglement, and it would be interesting to test whether they distinguish geometries with identical entanglement dynamics. Higher-dimensional and anisotropic black holes, including those with Kasner-like interiors, are natural targets: the HM equations of appendix~\ref{app:gen} apply directly, and these backgrounds would allow a comparison of how deeply HM surfaces and other observables penetrate the interior. Finally, entanglement growth after quantum quenches and in TFD-like states is accessible to many-body numerics and quantum simulators. Applying the present framework to such non-holographic data would address which effective bulk geometry, if any, they encode. Our results for superluminal data suggest that the training loss itself could then serve as a diagnostic of whether a geometric description exists within the class of metrics considered.

%
\acknowledgments
HSJ was supported by an appointment to the JRG Program at the APCTP through the Science and Technology Promotion Fund and Lottery Fund of the Korean Government. HSJ was also supported by the Korean Local Governments -- Gyeongsangbuk-do Province and Pohang City.
KYK was supported by the National Research Foundation of Korea (NRF) grant funded by the Korea government (MSIT) (RS-2026-25477372, RS-2025-02311201, RS-2025-02307394, RS-2024-00445164), and by the Creation of the Quantum Information Science R$\&$D Ecosystem (Grant No. 2022M3H3A106307411, RS-2023-NR068116). KYK was also supported by GIST research fund [Future leading Specialized Research Project, 2026 and the ANCHOR program through the Gwangju ANCHOR Center, funded by the Ministry of Education (MOE) and the Jeonnam-Gwangju Special Metropolitan City, Republic of Korea (2026-ANCHOR-05-001)].
Claude Opus 5.5 (Anthropic) and GPT-6 Sol (OpenAI) were used to assist in verifying the authors’ own calculations and results and in editing the manuscript; the authors take full responsibility for the content of this publication.
All authors contributed equally to this paper and should be considered as co-first authors.

\appendix
%
\section{Entanglement growth of half-lines in the CFT\texorpdfstring{$_2$}{2} thermofield double}\label{app:cft}

In this appendix we derive~\eqref{eq:btzS} directly in a two-dimensional CFT with central charge $c$, following~\cite{Calabrese:2004eu,Calabrese:2005in,Hartman:2013qma}.

The entanglement entropy is obtained from the R\'enyi entropies, $S_A=\lim_{n\to1}\frac{1}{1-n}\log\mathrm{Tr}\rho_A^n$.
For a region whose boundary consists of points, $\mathrm{Tr}\rho_A^n$ is the correlation function of twist operators $\sigma_n$, $\tilde\sigma_n$ inserted at the endpoints of $A$ in the $n$-fold replicated theory~\cite{Holzhey:1994we,Calabrese:2004eu}.
The twist operators are primaries with conformal weights
\begin{equation}
h_n=\bar h_n=\frac{c}{24}\left(n-\frac1n\right),\qquad \Delta_n\equiv h_n+\bar h_n=\frac{c}{12}\left(n-\frac1n\right).
\end{equation}
For $A=A_L\cup A_R$ with $A_{L,R}=\{x>0\}$ on the two copies, the only endpoints are $x=0$ on the left and on the right, so
\begin{equation}\label{eq:twist2pt}
\mathrm{Tr}\rho_A^n=c_n\,\epsilon^{2\Delta_n}\,\big\langle\mathrm{TFD}(t)\big|\,\sigma_n(0)_L\,\tilde\sigma_n(0)_R\,\big|\mathrm{TFD}(t)\big\rangle\,,
\end{equation}
where $\epsilon$ is the UV cutoff and $c_n$ a non-universal constant with $c_1=1$.

The TFD state is prepared by a Euclidean path integral over an interval of length $\beta/2$, and correlation functions of operators on the two sides equal thermal correlation functions on the cylinder of circumference $\beta$, with the left operators displaced by half the thermal circle~\cite{Maldacena:2001kr}.
With both copies evolved forward, $t_L=t_R=t$, the two twist operators sit at the same spatial position and are separated along the Euclidean time direction by the complex amount
\begin{equation}
\tau_{LR}=\frac{\beta}{2}+2it\,.
\end{equation}
The conformal map $\zeta=e^{2\pi w/\beta}$, with $w=x+i\tau$, takes the cylinder to the plane, and the two-point function of a primary of weights $(h,\bar h)$ becomes
\begin{equation}
\langle\phi(w_1,\bar w_1)\phi(w_2,\bar w_2)\rangle_\beta=\left[\frac{\beta}{\pi}\sinh\frac{\pi w_{12}}{\beta}\right]^{-2h}\left[\frac{\beta}{\pi}\sinh\frac{\pi\bar w_{12}}{\beta}\right]^{-2\bar h}.
\end{equation}
For $w_{12}=i\tau_{LR}$ and $\bar w_{12}=-i\tau_{LR}$ with $h=\bar h$, this gives $\big[\frac{\beta}{\pi}\sin\frac{\pi\tau_{LR}}{\beta}\big]^{-2\Delta}$, and using $\sin\big(\frac{\pi}{2}+\frac{2\pi i t}{\beta}\big)=\cosh\frac{2\pi t}{\beta}$,
\begin{equation}
\mathrm{Tr}\rho_A^n=c_n\left[\frac{\beta}{\pi\epsilon}\cosh\frac{2\pi t}{\beta}\right]^{-2\Delta_n}.
\end{equation}

Since $2\Delta_n/(n-1)\to c/3$ as $n\to1$, we obtain
\begin{equation}\label{eq:cftS}
S_A(t)=\frac{c}{3}\log\left(\frac{\beta}{\pi\epsilon}\cosh\frac{2\pi t}{\beta}\right)+c_1'\,,
\end{equation}
with $c_1'$ a non-universal constant.
Equation~\eqref{eq:cftS} is universal: it relies only on conformal invariance and on the dimension of the twist operators, and it holds at any central charge.
At $t=0$ it reduces to $S_A=\frac{c}{3}\log\frac{\beta}{\pi\epsilon}$, the entanglement entropy between two half-lines across the TFD, i.e.\ twice the familiar $\frac{c}{6}\log\frac{\beta}{\pi\epsilon}$ per entangling point.
At early times $S_A(t)-S_A(0)\simeq\frac{2\pi^2c}{3\beta^2}t^2$, while at late times $S_A(t)\simeq\frac{2\pi c}{3\beta}\,t=2s\,t$, with $s=\pi c/(3\beta)$ the thermal entropy density of a CFT$_2$; hence $\vE=1$.
With $c=3/(2\GN)$ and $\beta=2\pi\zh$, eq.~\eqref{eq:cftS} coincides with the holographic result~\eqref{eq:btzS}; the non-universal constant $c_1'$ vanishes for the holographic cutoff $z=\epsilon$.%

%
\section{Hartman--Maldacena surfaces in general dimensions}\label{app:gen}

We extend the Hartman--Maldacena (HM) surface formalism of section~\ref{sec:hm} to a $(d+1)$-dimensional planar black hole with metric
\begin{equation}\label{eq:gen_metric}
\dd s^2=\frac{1}{z^2}\left(-f(z)e^{-\chi(z)}\,\dd t^2 +\frac{\dd z^2}{f(z)}+\sum_{i=1}^{d-1}\dd x_i^2\right).
\end{equation}
The AdS boundary is at $z=0$, with $f(0)=1$ and $\chi(0)=0$, and the outer horizon is at $z=\zh$, where $f(\zh)=0$. Its surface gravity and entropy density are
\begin{equation}\label{eq:gen_thermo}
\kappa_h=2\pi T=-\frac12 f'(\zh)e^{-\chi(\zh)/2}, \qquad s=\frac{1}{4\GN\zh^{d-1}},
\end{equation}
where we assume a non-extremal outer horizon, $f'(\zh)<0$. We take $A_L$ and $A_R$ to be the half-spaces $x_1>0$ on the two boundaries. The HM surface is at $x_1=0$ and is extended along $x_2,\ldots,x_{d-1}$. Denote the regulated transverse volume by $\Omega_{d-2}$ and the HM area by $\cA$. We use the area density $a\equiv\cA/\Omega_{d-2}$, so that $S=\Omega_{d-2}a/(4\GN)$.

Parametrising one half of the surface by $t(z)$, its area density and conserved energy are
\begin{align}
a&=2\int\dd z\;\frac{1}{z^{d-1}} \sqrt{\frac{1}{f}-fe^{-\chi}\,t'(z)^2}, \label{eq:gen_length}\\
E&=\frac{fe^{-\chi}t'(z)}{z^{d-1} \sqrt{f^{-1}-fe^{-\chi}t'(z)^2}}. \label{eq:gen_E}
\end{align}
Here and below a prime on a metric function means a $z$-derivative; for $z(\lambda)$ and $V(\lambda)$, primes denote $\lambda$-derivatives. Introducing the effective potential
\begin{equation}\label{eq:gen_potential}
\Veff_d(z)\equiv-\frac{f(z)e^{-\chi(z)}}{z^{2(d-1)}},
\end{equation}
we find
\begin{equation}\label{eq:gen_tprime}
t'(z)=\frac{E e^{\chi/2}}{f\sqrt{E^2-\Veff_d(z)}}, \qquad E^2=\Veff_d(\zt).
\end{equation}
The turning point $\zt$ lies behind the horizon. The area density and symmetric boundary time are therefore
\begin{align}
a(E)&=2\int_{\epsilon}^{\zt}\dd z\; \frac{e^{-\chi/2}}{z^{2(d-1)}\sqrt{E^2-\Veff_d(z)}}, \label{eq:gen_AE}\\
t(E)&=-\PV\!\int_0^{\zt}\dd z\; \frac{E e^{\chi/2}}{f\sqrt{E^2-\Veff_d(z)}}. \label{eq:gen_tE}
\end{align}
The Hamilton--Jacobi relation generalizes \eqref{eq:HJ} to
\begin{equation}\label{eq:gen_HJ}
\frac{\dd a}{\dd t}=2E,\qquad \frac{\dd S}{\dd t}=\frac{\Omega_{d-2}}{2\GN}E.
\end{equation}
For a potential with an interior maximum, the critical surface $z=\zs$ satisfies
\begin{equation}\label{eq:gen_critical}
\Veff_d'(\zs)=0, \qquad \zs f'(\zs)-\zs f(\zs)\chi'(\zs) =2(d-1)f(\zs).
\end{equation}
At late times $E\to E_c\equiv\sqrt{\Veff_d(\zs)}$, and the entanglement velocity, defined by $\dd S/\dd t=2\Omega_{d-2}s\vE$, is
\begin{equation}\label{eq:gen_vE}
\vE=\zh^{d-1}E_c =\left(\frac{\zh}{\zs}\right)^{d-1} \sqrt{-f(\zs)e^{-\chi(\zs)}}.
\end{equation}

In the Kruskal-like coordinate $V$ defined in~\eqref{eq:Vcoord}, the metric becomes
\begin{equation}\label{eq:gen_metricV}
\dd s^2=\frac{1}{z^2}\left( -\frac{fe^{-\chi}}{\kappa_h^2 V^2}\,\dd V^2 -\frac{2e^{-\chi/2}}{\kappa_h V}\,\dd V\dd z +\sum_{i=1}^{d-1}\dd x_i^2\right).
\end{equation}
With $z=z(\lambda)$ and $V=V(\lambda)$, the area per unit transverse volume is $a=\int\dd\lambda\,\cL_d$, where
\begin{equation}\label{eq:gen_L}
\cL_d=\frac{1}{z^{d-1}}\left( -\frac{fe^{-\chi}}{\kappa_h^2 V^2}\,V'^2 -\frac{2e^{-\chi/2}}{\kappa_h V}\,V'z' \right)^{1/2}.
\end{equation}
Choosing the gauge $\cL_d^2=1$ as in section~\ref{sec:differential}, an independent second-order equation is
\begin{equation}\label{eq:gen_Veq}
V''=\left[1+\frac{e^{-\chi/2}}{2\kappa_h} \left(f'-f\chi'-\frac{2(d-1)}{z}f\right)\right] \frac{V'^2}{V}.
\end{equation}
The Killing vector $\partial_t=\kappa_h V\partial_V$ gives
\begin{equation}\label{eq:gen_EV}
E=-\frac{1}{z^{2(d-1)}}\left( \frac{fe^{-\chi}}{\kappa_h}\frac{V'}{V} +e^{-\chi/2}z'\right).
\end{equation}
Together with $\cL_d^2=1$, this implies
\begin{align}
z'^2&=z^{2(d-1)}f+E^2z^{4(d-1)}e^{\chi} =z^{4(d-1)}e^{\chi}\big(E^2-\Veff_d(z)\big), \label{eq:gen_scattering}\\
z''&=\frac12\frac{\dd}{\dd z}\left[ z^{2(d-1)}f(z)+E^2z^{4(d-1)}e^{\chi(z)}\right], \label{eq:gen_zeq}\\
\big(\log|V|\big)'&= \frac{\kappa_h z^{2(d-1)}} {Ez^{2(d-1)}-e^{-\chi/2}z'}. \label{eq:gen_logV}
\end{align}
For $d>2$, the parameter $\lambda$ measures area per unit transverse volume rather than proper length in the original $(t,z)$ metric; hence $a=\lambda_R-\lambda_L$.

We launch the surface at $\lambda=0$ from $(V,z)=(0,\zh)$. Writing $V_{(1)}=V'(0)$ and $z_{(1)}=z'(0)$, a regular series solution of~\eqref{eq:gen_Veq} and~\eqref{eq:gen_zeq} is
\begin{align}
V(\lambda)&=V_{(1)}\lambda +\left(\frac{d-1}{\zh}-\frac{f''(\zh)}{2f'(\zh)} +\frac34\chi'(\zh)\right) z_{(1)}V_{(1)}\lambda^2+\mathcal{O}(\lambda^3), \label{eq:gen_Vseries}\\
z(\lambda)&=\zh+z_{(1)}\lambda +\left[\frac14\zh^{2(d-1)}f'(\zh) +\left(\frac{d-1}{\zh}+\frac14\chi'(\zh)\right) z_{(1)}^2\right]\lambda^2+\mathcal{O}(\lambda^3), \label{eq:gen_zseries}\\
z_{(1)}&=e^{\chi(\zh)/2}\zh^{2(d-1)}E. \label{eq:gen_z1}
\end{align}
The rescaling freedom $V\to cV$ sets $V_{(1)}=1$, leaving $E$ fixed by~\eqref{eq:gen_HJ} as the independent horizon datum. At equal radial cutoffs $z(\lambda_R)=z(\lambda_L)=\epsilon$, the boundary readout is
\begin{equation}\label{eq:gen_readout}
t=\frac{1}{2\kappa_h} \log\left|\frac{V(\lambda_R)}{V(\lambda_L)}\right|, \qquad S=\frac{\Omega_{d-2}}{4\GN} (\lambda_R-\lambda_L).
\end{equation}
For the standard asymptotic falloff $f(z)=1+\mathcal{O}(z^d)$ (allowing logarithmic corrections at this order), the UV subtraction is
\begin{equation}\label{eq:gen_subtraction}
a_{\rm div}(\epsilon)= \begin{cases} 2\log(1/\epsilon),&d=2,\\[2pt]
\displaystyle\frac{2}{d-2}\,\epsilon^{-(d-2)},&d>2. \end{cases} \qquad S_{\rm reg}=\frac{\Omega_{d-2}}{4\GN} \lim_{\epsilon\to0}\left[ \lambda_R-\lambda_L-a_{\rm div}(\epsilon)\right].
\end{equation}
For more general falloffs, such as $f'(0)\neq0$ in $d>2$, further divergent terms must be subtracted to define $S_{\rm reg}$. The difference $S(t)-S(0)$ is independent of these time-independent subtractions.

For $d>2$, the potential~\eqref{eq:gen_potential} is not the original Killing norm $g_{tt}=-fe^{-\chi}/z^2$. It is instead the Killing norm of the effective two-dimensional metric
\begin{equation}\label{eq:gen_effective_metric}
\dd s_2^2=\frac{1}{z^{2(d-1)}} \left(-fe^{-\chi}\,\dd t^2+\frac{\dd z^2}{f}\right).
\end{equation}
Defining its local surface gravity by
\begin{equation}\label{eq:gen_kappa}
\kappa_d(z)\equiv\frac12z^{2(d-1)}e^{\chi/2} \frac{\dd\Veff_d}{\dd z} =e^{-\chi/2}\left[ \frac{(d-1)f}{z}-\frac12f'+\frac12f\chi'\right],
\end{equation}
we have $\kappa_d(\zh)=\kappa_h$ and $\kappa_d(\zs)=0$. Equivalently, $\kappa_d^2=-\hat\nabla_a\Veff_d\hat\nabla^a\Veff_d/ (4\Veff_d)$, with $\hat\nabla$ the covariant derivative of~\eqref{eq:gen_effective_metric}, as in~\eqref{eq:kappa}. On a branch where $\Veff_d$ increases monotonically from $-\infty$ at the boundary to $E_c^2$ at the critical surface, \eqref{eq:gen_AE} and~\eqref{eq:gen_tE} become
\begin{align}
\frac{4\GN}{\Omega_{d-2}}S(E) &=\int_{-\infty}^{E^2}\frac{\dd\Veff_d} {\kappa_d(\Veff_d)\sqrt{E^2-\Veff_d}}, \label{eq:gen_abelS}\\
t(E)&=\frac{E}{2}\PV\!\int_{-\infty}^{E^2} \frac{\dd\Veff_d} {\Veff_d\,\kappa_d(\Veff_d)\sqrt{E^2-\Veff_d}}. \label{eq:gen_abelt}
\end{align}
The first integral has the same UV regulation as~\eqref{eq:gen_AE}. The HM dynamics thus depends on the single invariant function $\kappa_d(\Veff_d)$ of the effective metric. At $d=2$ and $\Omega_0=1$, these equations reduce to the main-text results~\eqref{eq:metric}, \eqref{eq:thermo}, \eqref{eq:lengthfunc}--\eqref{eq:vE}, \eqref{eq:metricV}--\eqref{eq:z1E}, \eqref{eq:tV}, \eqref{eq:kappa} and~\eqref{eq:abelform}.

\paragraph{Degeneracy.}
Since~\eqref{eq:gen_abelS} and~\eqref{eq:gen_abelt} depend on the metric only through $\kappa_d(\Veff_d)$, the degeneracy is present in any dimension: for any smooth monotonic $Z(z)$ with $Z(0)=0$, $Z'(0)=1$ and $Z(\zh)=\zh$,
\begin{equation}\label{eq:gen_degenerate}
\tilde f\big(Z(z)\big)=\left(\frac{z}{Z(z)}\right)^{2(d-1)}Z'(z)^2 f(z)\,,\qquad
\tilde\chi\big(Z(z)\big)=\chi(z)+2\log\frac{z^{2(d-1)}Z'(z)}{Z(z)^{2(d-1)}}
\end{equation}
leaves the effective metric~\eqref{eq:gen_effective_metric}, and hence $S(t)-S(0)$, unchanged, while the transverse metric $\sum_i\dd x_i^2/z^2$ changes.
At $d=2$, eq.~\eqref{eq:gen_degenerate} reduces to~\eqref{eq:degenerate}.

\providecommand{\href}[2]{#2}\begingroup\raggedright\endgroup

\end{document}